\documentclass[11pt]{article}
\pdfoutput=1
\usepackage{amsmath,amssymb,mathtools}
\usepackage{graphicx}
\usepackage{units}
\usepackage{url}
\usepackage{booktabs}
\usepackage{subfigure}
\usepackage{hyperref}

\usepackage[sc,small]{caption}
\usepackage{color}

\DeclareFontFamily{U}{mathx}{\hyphenchar\font45}
\DeclareFontShape{U}{mathx}{m}{n}{<-> mathx10}{}
\DeclareSymbolFont{mathx}{U}{mathx}{m}{n}
\DeclareMathAccent{\wb}{0}{mathx}{"73}

\newcommand{\mathscr}{\mathcal}

\newcommand{\be}{\begin{eqnarray}}
\newcommand{\ee}{\end{eqnarray}}
\newcommand{\bea}{\begin{eqnarray}}
\newcommand{\eea}{\end{eqnarray}}

\newcommand{\tht}{\vartheta}

\newcommand{\A}{{\mathcal A}}
\newcommand{\K}{{\mathcal K}}
\newcommand{\M}{{\mathcal M}}
\newcommand{\N}{{\mathcal N}}

\newcommand{\thba}[2]{\vartheta[\!\!\begin{array}{c}{\phantom{}\vspace{-0.5mm}\scriptstyle#1}%
                        \\[-1.8mm]{\scriptstyle #2}\end{array}\!\!]}

\newcommand{\ba}[2]{[\!\!\begin{array}{c}{\scriptstyle#1}%
                        \\[-1.6mm]{\scriptstyle #2}\end{array}\!\!]}                         
\newcommand{\beqn}{\begin{eqnarray}}
\newcommand{\eeqn}{\end{eqnarray}}

\newcommand{\ca}{{\cal A}}
\newcommand{\ck}{{\cal K}}

\newcommand{\cn}{{\cal N}}
\newcommand{\cm}{{\cal M}}

\newcommand{\ct}{{\cal T}}

\newcommand{\cv}{{\cal V}}

\newcommand{\ImT}{\tau}
\newcommand{\ImS}{\sigma}

\newcommand{\p}{\partial}

\newcommand{\dd}{\text{d}}

\renewcommand{\ss}[2]{\Big[\genfrac{}{}{0pt}{1}{#1}{#2}\Big]}

\newcommand{\nn}{ \nonumber }
\newcommand{\hf}{ { \frac{1}{2} } }
\newcommand{\thf}{ { \textstyle \frac{1}{2} } }

\newcommand{\vev}[1]{\langle #1 \rangle}

\newcommand{\reftab}[1]{table \ref{#1}}

\newcommand{\non}{\nonumber \\}

\DeclareMathOperator{\Cl}{\mathrm{Cl}}

\newcommand{\vt}{\vartheta}

\begin{document}

\hfill LMU-ASC 70/15, NITS-PHY-2015009\\

\begin{center}
{\bf\LARGE
One-loop Einstein-Hilbert term in minimally supersymmetric type IIB orientifolds}

\vspace{1.5cm}

{\large
{\bf Michael Haack$^{\dag}$, }
{\bf Jin U Kang$^{\star}$}
\vspace{1cm}

{\it
$^{\dag}$ 
 Arnold Sommerfeld Center for Theoretical Physics \\ 
Ludwig-Maximilians-Universit\"at M\"unchen \\ 
Theresienstrasse 37, 80333 M\"unchen, Germany\\ [5mm] 
$^{\star}$ 
Department of Physics, Nanjing University\\  22 Hankou Road, Nanjing 210093, PR.\ China\\
and\\
Abdus Salam International Centre for Theoretical Physics\\  Strada Costiera 11, Trieste 34014, Italy 
\\and\\
Department of Physics, Kim Il Sung University\\ RyongNam Dong, TaeSong District, Pyongyang, DPR.\ Korea 
 \\[5mm]
}
}

\end{center}
\vspace{2mm}

\begin{center}
{\bf Abstract}\\
\end{center}
We evaluate 
string one-loop contributions
to the Einstein-Hilbert term
 in toroidal minimally supersymmetric type IIB orientifolds
with D-branes. These have potential applications to the determination of 
quantum corrections to the moduli K\"ahler metric in these models.  
\clearpage
 \tableofcontents

\section{Introduction}

Toroidal orientifolds with minimal supersymmetry are valuable tools for string phenomenology (see e.g.\ \cite{Angelantonj:2002ct,Blumenhagen:2006ci} for reviews). On the one hand they are rich enough to be phenomenologically interesting (for example, having $\cn = 1$ supersymmetry), on the other hand, they are simple enough to be technically rather tractable. To make progress in string phenomenology one could try different routes. One could either try to find universal or at least generic features in string model building or one could try to understand particular models in great detail, either with the hope that the features of the model are representative at least for a certain class of string compactifications or with the aim to see what possibilities string theory offers within a particular model. Having this second approach in mind, this article continues the investigation of 1-loop corrections to the K\"ahler metric of toroidal $\cn = 1$ type II orientifolds with emphasis on the effect of $\cn = 1$ sectors, initiated in \cite{Berg:2011ij,Berg:2014ama}.  

Whereas \cite{Berg:2014ama} dealt with the 1-loop correction to the kinetic term of the moduli scalars in toroidal $\cn = 1$ type IIB orientifolds, here we are focusing on the 1-loop correction to the Einstein-Hilbert term in string frame. The two questions are, however, closely related as we will review in sec.\ \ref{sec:effective}. In order to determine the 1-loop correction to the K\"ahler metric of the scalar manifold, a knowledge of the quantum corrections to the Einstein-Hilbert term is indispensable. Therefore, with this interrelation in mind, our main example in this paper will be the $\mathbb{Z}_6'$ orientifold that was also the main example in \cite{Berg:2014ama}. 

Calculating 1-loop corrections to the Einstein-Hilbert term within string theory is not a new subject. In supersymmetric heterotic string compactifications such corrections are actually absent \cite{Antoniadis:1992sa,Kiritsis:1994ta}, whereas there are non-trivial corrections in four dimensional type I models with at most $\cn = 2$ supersymmetry. In the case of $\cn = 2$ models these were first calculated in \cite{Antoniadis:1996vw} (see also \cite{Antoniadis:2002tr}). The results were subsequently generalized to $\cn = 1$ orientifolds in \cite{Kohlprath:2003pu,Epple:2004ra}. However, whereas \cite{Kohlprath:2003pu} considered only non-compact type IIB $\mathbb{Z}_N$ orientifolds with odd $N$, \cite{Epple:2004ra} dealt with type IIA orientifolds with D6-branes at angles. 

In this paper we fill the missing gap and consider the 1-loop correction for a general compact and tadpole-free $\mathbb{Z}_N$ type IIB orientifold as enumerated in \cite{Aldazabal:1998mr}, i.e.\ $\mathbb{Z}_3, \mathbb{Z}_6, \mathbb{Z}_6', \mathbb{Z}_7, \mathbb{Z}_{12}$. Our general discussion is rather similar to the type IIA case treated in \cite{Epple:2004ra} but we also find differences. We would like to stress that the two cases are not T-dual to each other. Rather, under T-duality the type IIB orientifolds we are discussing here would be mapped to {\it asymmetric} type IIA orientifolds, see for instance the discussion in sec.\ 4 of \cite{Angelantonj:1999xf}.

We then apply our general discussion to two concrete examples, one with odd $N$ and one with even $N$, i.e.\ the $\mathbb{Z}_3$ and the $\mathbb{Z}_6'$ models. One reason to consider also a $\mathbb{Z}_N$ orientifold with odd $N$ was the claim of \cite{Kohlprath:2003pu} that the contributions arising from the annulus $\A$, M\"obius $\M$ and Klein bottle $\K$ vanish in the case of odd $N$ type IIB orientifolds, leaving only the torus contribution. We disagree with this and identify a possible source for the discrepancy. 

Apart from the conceptional differences to \cite{Epple:2004ra}, we also had to extend that work on a technical level. Applying our general discussion to the case of the $\mathbb{Z}_6'$ orientifold requires the evaluation of a new type of integral over the world-sheet parameter. These integrals arise for the annulus amplitude with one end on D5-branes and one on D9-branes as well as for the twisted Klein bottle. There were no analogous contributions in the examples considered in \cite{Epple:2004ra}. We evaluate this new type of integral in app.\ \ref{t-integral_n1}, following a similar calculation in \cite{Berg:2014ama}. 

Similar to \cite{Epple:2004ra}, we find that there are two different kinds of corrections to the Einstein-Hilbert term. Those arising from the $\cn = 2$ sectors of the orientifold are similar to the ones found in \cite{Antoniadis:1996vw}. They have a complicated dependence on the complex structure of the compactification space. However, in contrast to the purely $\cn = 2$ supersymmetric case discussed in \cite{Antoniadis:1996vw} some of these moduli dependent corrections in the $\cn = 1$ models do not vanish when decompactifying the internal directions. This was also observed in \cite{Epple:2004ra}. In addition to these complex structure dependent corrections, in $\cn = 1$ models there are also corrections arising from the $\cn = 1$ sectors of the orientifold. These are moduli independent numbers. 

The work of \cite{Antoniadis:2002tr,Kohlprath:2003pu,Epple:2004ra} was primarily motivated by attempts to find an embedding of the Dvali-Gabadadze-Porrati scenario \cite{Dvali:2000hr} into string theory. As we mentioned above, our main motivation is very different. We are aiming at a better understanding of the quantum corrections to the low energy effective action of the $\mathbb{Z}_6'$ model. On the way to determine the 1-loop correction to the K\"ahler metric on the moduli space the knowledge of the 1-loop correction to the Einstein-Hilbert term is a necessary step. The third and final task to completely determine the 1-loop correction to the K\"ahler metric requires a knowledge of the correct definition of the field variables at loop level. Examples of a field redefinition necessitated by quantum corrections can be found for instance in \cite{Antoniadis:1996vw,Antoniadis:2003sw,Blumenhagen:2007ip,Camara:2009uv,Grimm:2013bha}. We leave this task for future work. 

The paper is organized as follows. In sec.\ \ref{sec:effective} we review the necessity of a knowledge of the quantum corrections to the Einstein-Hilbert term when discussing quantum corrections to the moduli metric in the low energy effective action. In sec.\ \ref{1loop-2pt} we discuss 1-loop corrections to the Planck mass in $\cn = 1$ type IIB toroidal orientifolds, focusing on general statements and formulas, i.e.\ without specializing to a particular $\mathbb{Z}_N$ model. In the following two sections, these formulas are then evaluated in the case of two concrete cases, the $\mathbb{Z}_3$ and the $\mathbb{Z}_6'$ models (in sec.\ \ref{z3} and sec.\ \ref{z6}, respectively). We refer readers, who are mainly interested in the final results, directly to equations \eqref{delta_Z3_final} and \eqref{finaldE}, which give the 1-loop correction to the Einstein-Hilbert term, i.e.\ $\delta E$ as defined in \eqref{kinetic_term}, for $\mathbb{Z}_3$ and $\mathbb{Z}_6'$, respectively ($E_2$ appearing in \eqref{finaldE} is the non-holomorphic Eisentein series, defined in \eqref{eisenstein}). Finally we conclude in sec.\ \ref{sec:concl}. Moreover, we collect some technical details in the appendix. More concretely, in app.\ \ref{useful} we give a few useful formulas, in app.\ \ref{partfun} we exemplarily give the full partition function of the $\mathbb{Z}_6'$ orientifold in order to illustrate the compact formulas of the main text. Then in app.\ \ref{t-integral} we gather the details of two integrals that are needed in the main text. The first one, given in \ref{t-integral_n1}, is relevant for the contributions from $\cn = 1$ sectors and is new to our knowledge.


\section{Effective field theory}
\label{sec:effective}

In this section we review how the quantum corrections to the Einstein-Hilbert term influence the form of the low energy effective action of string compactifications. This discussion heavily draws from sec.\ 2 in \cite{Berg:2014ama}. 

As discussed there, in order to determine the quantum corrected K\"ahler metric on the moduli manifold in the Einstein frame, one has to deal with two complications, in addition to calculating the direct 1-loop correction to the metric in the string frame (which was the focus of \cite{Berg:2014ama}): One needs to know the quantum corrections to the Einstein-Hilbert term in the string frame (which is the focus of the present paper) and the quantum corrected definition of the K\"ahler variables on moduli space (which we leave for future work). 

In order to exemplify these issues let us concentrate, following \cite{Berg:2014ama}, on the K\"ahler modulus of the third torus in a toroidal orientifold model, which we will denote by $\tau$. Its tree level definition is denoted by $\tau^{(0)}$. Now the quantum corrected kinetic term of $\tau^{(0)}$ coupled to gravity in string frame and up to 1-loop order is given by
\be \label{kinetic_term}
S_4 =  \frac{1}{\kappa_4^2} \int d^4 x \sqrt{-g} \left[ \left( e^{-2 \Phi_4} + \delta E \right) {1 \over 2} R  + \left( \widetilde G^{(0)} + \widetilde G^{(1)} \right) \partial_\mu \ImT^{(0)} \partial^\mu \ImT^{(0)} \right]+ \ldots\ ,
\ee
where $\delta E$ denotes the correction to the Einstein-Hilbert term, including tree level $\alpha'$-corrections and corrections from 1-loop, $\widetilde G^{(0)}$ stands for the tree level metric (including $\alpha'$ corrections \cite{Becker:2002nn}) and $\widetilde G^{(1)}$ denotes the contributions to the string frame metric arising at 1-loop level. Moreover, 
\be \label{kappa4}
\kappa_4^{-2} = (2 \pi \sqrt{\alpha'})^6 \kappa_{10}^{-2}=(\pi \alpha')^{-1}
\ee
and
\be \label{Phi}
e^{-2 \Phi_4} \equiv e^{-2 \Phi_{10}} t_1 t_2 t_3 = \sqrt{\ImS^{(0)} \ImT_1^{(0)} \ImT_2^{(0)} \ImT_3^{(0)}} \ ,
\ee
where $e^{-2 \Phi_{10}}$ is the ten-dimensional dilaton and
\be \label{Tt}
\ImS^{(0)}  =  e^{-\Phi_{10}} t_1 t_2 t_3\quad , \quad 
\ImT_i^{(0)} = e^{-\Phi_{10}} t_i\ .
\ee
Here the $t_i$ are the (dimensionless) torus volumes measured with the string frame metric.\footnote{More concretely, $t_i = \frac{V_i}{4 \pi^2 \alpha'}$, where $V_i$ are the torus volumes.} As mentioned above, when we talk about $\tau$ without a subscript we always have $\tau_3$ in mind. 

As discussed in \cite{Berg:2014ama}, the definition of the K\"ahler variables in general gets quantum corrected, i.e.\ one has
\be \label{deltaImT}
\ImT = \ImT^{(0)}  + \delta \tau\ ,
\ee
where $\delta \tau$ is a moduli dependent function. In the case of an $\cn = 2$ supersymmetric compactification on $K_3 \times T^2$ at an orbifold point this function $\delta \tau$ was determined at 1-loop level in \cite{Antoniadis:1996vw}. In general there might also be corrections from the disk level, in particular in the presence of fluxes and open string scalars \cite{Antoniadis:1996vw, Blumenhagen:2007ip, Camara:2009uv,Grimm:2013bha}. 

Starting from \eqref{kinetic_term} and performing a Weyl transformation to go to the Einstein frame, it was shown in \cite{Berg:2014ama} that the quantum correction to the metric of the quantum corrected K\"ahler modulus $T$ (with imaginary part $\tau$), is given, up to 1-loop order, by\footnote{In deriving this result, some doubly suppressed terms were neglected, i.e.\ those which are suppressed both in the large volume $\cv^{-1}$ and the small string coupling $g_s$. See \cite{Berg:2014ama} for more details.} 
\be
G^{(1)}_{T \bar T}(T) &=& e^{2 \Phi_4} \widetilde G^{(1)}(\ImT) + 12 \left(\frac{\partial \Phi_4}{\partial \ImT^{(0)}}\right)^2  \delta E e^{2 \Phi_4} + 6 \frac{\partial \Phi_4}{\partial \ImT^{(0)}} \frac{\partial \delta E}{\partial \ImT^{(0)}} e^{2 \Phi_4} \nn \\
&& - \delta E e^{4 \Phi_4} \widetilde G^{(0)}(\ImT) + \frac{1}{2 \ImT^3} \delta \tau - \frac{1}{2 \ImT^2} \frac{\partial  \delta \tau}{\partial \ImT} + \ldots\ .  \label{finalmetric} 
\ee
Obviously, a knowledge of the quantum correction to the Einstein-Hilbert term $\delta E$ is crucial for a complete understanding of the quantum corrected K\"ahler metric. Its determination for toroidal $\cn = 1$ type IIB orientifolds is the subject of the following sections. 


\section{Graviton 1-loop 2-point function, general analysis} 
\label{1loop-2pt}

In this section we derive some general formulas needed for computing the 1-loop correction to the Planck mass in ${\cal N} = 1$ type IIB toroidal orientifolds. These will be applied in secs.\ \ref{z3} and \ref{z6} to the concrete $\mathbb{Z}_3$ and $\mathbb{Z}_6'$ models. At the beginning we follow closely the presentation in \cite{Epple:2004ra}. In the concrete evaluation of the resulting formulas, our approaches differ (also from the approach pursued in \cite{Kohlprath:2003pu}). We found it most efficient to perform the spin-structure sum early on and go to the tree channel only at the very end. Moreover, in contrast to \cite{Epple:2004ra}, applying the general formulas to the $\mathbb{Z}_6'$ orientifold required evaluating a new type of integral that we perform in app.\ \ref{t-integral_n1}.

Starting point is the amplitude of two gravitons (with momenta $p_i$ and polarization tensors $\varepsilon_i$)
\be \label{gravitonamplitude}
\big\langle\!\! \big\langle V_{g}(p_1,\varepsilon_1) V_{g}(p_2,\varepsilon_2) \big\rangle\!\!  \big\rangle = \sum_{\sigma \in \{ \ct, \K, \A, \M \}} \big\langle\!\! \big\langle V_{g}(p_1,\varepsilon_1) V_{g}(p_2,\varepsilon_2) \big\rangle\!\!  \big\rangle_\sigma\ ,
\ee
where the vertex operators are given by\footnote{We follow the conventions of \cite{Polchinski:1998rr} which slightly differ from the ones used in \cite{Epple:2004ra}.}
\be \label{vop}
V_{g}(p, \varepsilon) = -\frac{2 g_c}{\alpha'} \varepsilon_{\mu \nu} \left(i \partial X^\mu + \tfrac{\alpha'}{2} p \cdot \psi\, \psi^\mu \right) \left(i \bar \partial X^\nu + \tfrac{\alpha'}{2} p \cdot \tilde \psi\, \tilde \psi^\nu \right) e^{i p\cdot X}
\ee
with $\varepsilon_{\mu \nu} \varepsilon^{\mu \nu}=1$, and one has to sum up the contribution of all 1-loop surfaces $\sigma$. Using the on-shell, transversality and tracelessness conditions 
\be
p_1^2=p_2^2=p_1\cdot p_2 = p_{1\mu} \varepsilon_1^{\mu \nu} = p_{2\mu} \varepsilon_2^{\mu \nu} = \eta_{\mu \nu} \varepsilon_1^{\mu \nu} = \eta_{\mu \nu} \varepsilon_2^{\mu \nu}=0\ , \label{onshell}
\ee 
the amplitude \eqref{gravitonamplitude} has to be proportional to the only remaining contraction, i.e.
\be 
\big\langle\!\! \big\langle V_{g}(p_1,\varepsilon_1) V_{g}(p_2,\varepsilon_2) \big\rangle\!\!  \big\rangle =  A\, i V_4 g_c^2 p_2^\mu \varepsilon_{1 \mu \nu} \eta^{\nu \lambda} \varepsilon_{2 \lambda \rho} p_1^\rho + {\cal O}(p^4)\ . \label{defineA}
\ee
This defines the quantity $A$. $V_4$ is the regularized volume of the four-dimensional spacetime. Strictly speaking the contraction appearing in \eqref{defineA} is also vanishing due to momentum conservation and transversality. However, it was argued in \cite{Antoniadis:1996vw,Antoniadis:2002tr} that reading off the coefficient $A$ of the kinematically vanishing factor $p_2^\mu \varepsilon_{1 \mu \nu} \eta^{\nu \lambda} \varepsilon_{2 \lambda \rho} p_1^\rho$ in \eqref{defineA} gives the same result as a more rigorous calculation using a 3-point function. We assume that this still holds in the case of the $\cn = 1$ models under consideration here (\cite{Antoniadis:1996vw,Antoniadis:2002tr} considered a model with $\cn = 2$ supersymmetry).

In order to translate \eqref{defineA} into a correction to the four-dimensional Planck mass, we have to compare it to the relevant term in the action which leads to the linearized Einstein equations for a metric fluctuation fulfilling the conditions \eqref{onshell}. From eq.\ (6.9) in \cite{Carroll:1997ar}, for instance, we read off 
\be
S = \frac{M_P^2}{2} \int d^4x\, \Big( -\frac12 h_{\mu \nu, \rho} h^{\nu \rho, \mu} \Big)\ ,
\ee
where 
\be
G_{\mu \nu} = \eta_{\mu \nu} + h_{\mu \nu}\ ,
\ee
for a symmetric fluctuation $h_{\mu \nu}$. Note that the relation between $h_{\mu \nu}$ and the polarisation tensor $\varepsilon_{\mu \nu}$ appearing in the vertex operator \eqref{vop} is given by (in momentum space)
\be \label{hepsilon}
h_{\mu \nu} = - 4 \pi g_c \varepsilon_{\mu \nu} e^{i p\cdot X}\ ,
\ee
cf.\ (3.7.11) in \cite{Polchinski:1998rq}. Using the notation of \eqref{kinetic_term}, we have
\be
M_P^2 = \frac{1}{\kappa_4^2} \left( e^{-2 \Phi_4} + \delta E \right)\ ,
\ee
where $\kappa_4^{-2}$ was given in \eqref{kappa4}. Thus we should compare \eqref{defineA} with 
\be
-\frac14 \kappa_4^{-2} \int d^4x\, \delta E\, h_{\mu \nu, \rho} h^{\nu \rho, \mu}\ . \label{compare1loop}
\ee
In order to do so, in \eqref{defineA} we make the substitutions (cf.\ \eqref{hepsilon})
\be
V_4 \rightarrow \int d^4x \quad , \quad i g_c^2\, p_2^\mu \varepsilon_{1 \mu \nu} \eta^{\nu \lambda} \varepsilon_{2 \lambda \rho} p_1^\rho \rightarrow -\frac{1}{16 \pi^2} \cdot \frac12 \cdot h_{\mu \nu, \rho} h^{\nu \rho, \mu}\ ,
\ee
where the factor $1/2$ is a symmetry factor for identical fields and the factor of $i$ on the left hand side is the familiar factor for Lorentzian S-matrix elements. Comparing the resulting expression with \eqref{compare1loop}, we obtain
\be
\delta E = \frac{\kappa_4^2}{8 \pi^2} A = \frac{\alpha'}{8 \pi} A\ . \label{deltaE} 
\ee
Thus, the remaining task is to obtain an explicit expression for \eqref{defineA} in order to determine $A$. 

The amplitude gets contributions from all 1-loop surfaces, i.e.\ $\ct, \K, \A$ and $\M$. The torus contribution could be calculated via world-sheet methods, cf.\ \cite{Kohlprath:2003pu}, but we just read it off from eq.\ (5.3) in \cite{Antoniadis:1997eg}. Including also the $\alpha'$-correction to the Planck-mass from the sphere it gives
\be \label{deltaETS}
(\delta E)_{S_2+\ct}  = \frac{\chi}{(2 \pi)^3} \Big( 2 \zeta (3) \frac{e^{-2 \Phi_4}}{{\cal V}} + \frac{\pi^2}{3} \Big)\ ,
\ee
where ${\cal V}$ is the overall volume (in units of $(2 \pi \sqrt{\alpha'})^6$) and, due to the orientifold projection, we added a factor of $1/2$ to the torus contribution of eq.\ (5.3) in \cite{Antoniadis:1997eg}. Thus, in the following we can concentrate on the contribution from $\K, \A$ and $\M$.\footnote{It would be interesting to confirm that there are no contributions from the disk in $\cn = 1$ models, following the suggestion at the end of sec.\ 3 in \cite{Antoniadis:2002tr}.} 

We closely follow the calculation in \cite{Epple:2004ra}. Neglecting the momentum conservation delta function (arising from the bosonic zero mode integration) we have
\be
A_\sigma &=& - \frac{1}{8N (4 \pi^2 \alpha')^2}\, \sum_{s=\text{even}} \int_0^\infty \frac{d t}{t^3}
\, \sum_{k=0}^{N-1}\,Z_\sigma^{(k)}(\tau_\sigma, s) \int_\sigma d^2 \nu_1 \int_\sigma d^2 \nu_2 \\
&& ~~~~~~~~~~~~~~~~~~~~~~~ \Big(  \vev{ \bar \p X_1 \bar \p X_2 }_\sigma
 (\vev{ \psi_2 \psi_1 }_\sigma^s)^2
 +
 \vev{ \p X_1 \bar \p X_2 }_\sigma
 (\vev{ \psi_2 \tilde \psi_1 }_\sigma^s)^2 \nonumber \\
 && ~~~~~~~~~~~~~~~~~~~~~~~~ +
 \vev{ \bar \p X_1 \p X_2 }_\sigma
 (\vev{ \tilde \psi_2 \psi_1 }_\sigma^s)^2
 +
 \vev{ \p X_1 \p X_2 }_\sigma
 (\vev{ \tilde \psi_2 \tilde \psi_1 }_\sigma^s)^2 \Big)\ , \label{Asigma}
\ee
where $\sigma$ stands for the different world-sheet topologies $\K, \A$ and $\M$, with world-sheet parameters $\tau_\K = 2 i t$, $\tau_\A = \frac{i t}{2}$, $\tau_\M = \frac{1}{2}+\frac{i t}{2}$, and $Z_\sigma^{(k)}(\tau_\sigma, s)$ is the contribution to the partition function from the $k$-twisted sector. We will discuss it in more detail below, cf.\ \eqref{Z_sigma}. The spin structure sum only runs over the even spin structures $s$. Note, that there is no contribution to $A_\sigma$ from eight fermion terms. From \eqref{vop} these come with four powers of momenta and there are no poles in the $\nu$ integrals which could reduce the order in momenta (cf.\ sec.\ 3.4 in \cite{Berg:2014ama}). 

We now use (see for instance \cite{Epple:2004ra})
\be
(\vev{ \psi_2 (\nu) \psi_1 (0)}_\sigma^s)^2 = -\partial_\nu^2 \ln \vartheta_1(\nu , \tau) + \left. \partial_v^2 \frac{\vartheta_s(v, \tau)}{\vartheta_s(0, \tau)} \right|_{v=0}\ , \label{szegosquare}
\ee
i.e.\ it is the sum of a spin structure independent term with a spin structure dependent term. As argued in \cite{Kohlprath:2003pu,Epple:2004ra}, the contribution to $A_\sigma$ involving the first term in \eqref{szegosquare} (the spin structure independent term) does not survive the sum over spin structures in the supersymmetric case. On the other hand, the spin structure dependent (second) term does not depend on the vertex operator position and, thus, can be taken out of the $\nu$ integrals. Moreover, given that it does not depend on the vertex operator position, it is the same for $(\vev{ \psi_2 \psi_1 }_\sigma^s)^2$, $(\vev{ \psi_2 \tilde  \psi_1 }_\sigma^s)^2$, $(\vev{ \tilde \psi_2 \psi_1 }_\sigma^s)^2$ and $(\vev{ \tilde \psi_2 \tilde \psi_1 }_\sigma^s)^2$. Our conventions for the world-sheet fermions lead to relative minus signs between the contributions in \eqref{Asigma} arising from $(\vev{ \psi_2 \psi_1 }_\sigma^s)^2$ and $(\vev{ \tilde \psi_2 \tilde \psi_1 }_\sigma^s)^2$ on the one hand and $(\vev{ \psi_2 \tilde  \psi_1 }_\sigma^s)^2$ and $(\vev{ \tilde \psi_2 \psi_1 }_\sigma^s)^2$ on the other hand (cf.\ app.\ D in \cite{Berg:2014ama}). The resulting $\nu$ integral can then be solved using \cite{Antoniadis:1996vw}
\be
\int_\sigma d^2 \nu_1 \int_\sigma d^2 \nu_2 \Big( \vev{ \bar \p X_1 \bar \p X_2 }_\sigma - \vev{ \p X_1 \bar \p X_2 }_\sigma - \vev{ \bar \p X_1 \p X_2 }_\sigma + \vev{ \p X_1 \p X_2 }_\sigma \Big) = \frac{\alpha' \pi\, {\rm Im}(\tau_\sigma)}{2}\ .
\ee
Taking into account \eqref{deltaE}, we end up with\footnote{This is the analog of eqs.\ (2.20-22) in \cite{Epple:2004ra}. Our result differs slightly in the overall factor, due to \eqref{hepsilon}. Note that we do agree on the overall sign despite appearance. The sum over spin structures involves an extra minus sign in \cite{Epple:2004ra}, as can be seen for instance in eq.\ (3.3) therein.\label{signdiff}}
\be
(\delta E)_\sigma &=& \left. - \frac{\alpha'}{8 \pi} \frac{1}{8N (4 \pi^2 \alpha')^2}\,\partial_v^2\,    \sum_{s=\text{even}} \int_0^\infty \frac{d t}{t^3}
\, \sum_{k=0}^{N-1}\,Z_\sigma^{(k)}(\tau_\sigma, s)\, \frac{\vt_s(v,\tau_\sigma)}{\vt_s(0,\tau_\sigma)} \frac{\alpha' \pi\, {\rm Im}(\tau_\sigma)}{2}\right|_{v=0}  \\
& =& \left. - \frac{(\alpha')^2}{8 \pi}  \frac{1}{8N (4 \pi^2 \alpha')^2}\,\int_0^\infty \frac{d t}{t^3}\, \frac{\pi\, {\rm Im}(\tau_\sigma)}{2}
 \sum_{k=0}^{N-1}\, \partial_v^2 \sum_{s=\text{even}} 
\,Z_\sigma^{(k)}(\tau_\sigma, s)\, \frac{\vt_s(v,\tau_\sigma)}{\vt_s(0,\tau_\sigma)} \right|_{v=0} \\
& =& - \frac{(\alpha')^2}{8 \pi}  \frac{1}{8N (4 \pi^2 \alpha')^2}\,\int_0^\infty \frac{d t}{t^3}\, \frac{\pi\, {\rm Im}(\tau_\sigma)}{2}
 \sum_{k=0}^{N-1}\,  \sum_{s=\text{even}} 
\,Z_\sigma^{(k)}(\tau_\sigma, s)\, \frac{\vt_s''(0,\tau_\sigma)}{\vt_s(0,\tau_\sigma)}\ . \label{1pt} 
\ee
The partition functions can compactly be written as \cite{Aldazabal:1998mr,Berg:2014ama}
\be
Z_\sigma^{(k)}(\tau_\sigma, s)=(-2\pi)\,{\rm CP}_\sigma\, \tilde{\chi}_\sigma (-2 \sin(\pi \gamma_3)) \left(\prod_{j=1}^2 f(\gamma_j) \right)
   Z_s^\vt\big(\gamma_i, h_i,g_i \big)  \label{Z_sigma}
\ee
with $Z_s^\vt(\gamma_i, h_i, g_i)$ being the $\vt$-dependent part of the partition function given by 
\be
Z_s^\vt(\gamma_i, h_i,g_i) =  \eta_{\alpha \beta} \frac{\vt\ss{\alpha}{\beta} \vt\ss{\alpha+h_1}{\beta+\gamma_1+g_1}
\vt\ss{\alpha+h_2}{\beta+\gamma_2+g_2}\vt\ss{\alpha}{\beta+\gamma_3}}{\vt'\ss{\frac{1}{2}}{\frac{1}{2}}
\vt\ss{\frac{1}{2}+h_1}{\frac{1}{2}+\gamma_1+g_1}\vt\ss{\frac{1}{2}+h_2}{\frac{1}{2}+\gamma_2+g_2}
\vt\ss{\frac{1}{2}}{\frac{1}{2}+\gamma_3}}  \label{Z_s^vt}\ ,
\ee
where the relation between $s$ and $(\alpha,\beta)$ can be found in table \ref{table:spinstructures}. In \eqref{Z_sigma} ${\rm CP}_\sigma$  stands for the corresponding Chan-Paton factor of the open-string world-sheets and ${\rm CP}=1$ for the Klein bottle. Formula \eqref{Z_sigma} holds for all tadpole-free $\mathbb{Z}_N$ type IIB orientifolds discussed in  \cite{Aldazabal:1998mr}, i.e.\ $\mathbb{Z}_3, \mathbb{Z}_6, \mathbb{Z}_6', \mathbb{Z}_7, \mathbb{Z}_{12}$. The concrete forms of ${\rm CP}_\sigma$, $\tilde{\chi}_\sigma$, $\gamma_i$, $f$, $h_i$ and $g_i$ can be found in \reftab{tab:details}. The models with even $N$ have D5-branes wrapped around the third torus leading to the distinction of $\gamma_3$ in \eqref{Z_sigma}. On the other hand, models with odd $N$ do not have any D5-branes (and thus, no amplitudes $\ca_{55}, \ca_{95}, \cm_{5}$) and only untwisted strings run in the Klein bottle (i.e.\ there is no amplitude $\ck_t$).  
%
\begin{table}[t]
\renewcommand{\arraystretch}{1.3}
\begin{center}
\begin{tabular}{ccccc}
\toprule
$s$ & 1 & 2 & 3 & 4 \\
\midrule
$\ss{\alpha}{\beta}$ & $\ss{1/2}{1/2}$ & $\ss{1/2}{0}$ & $\ss{0}{0}$ & $\ss{0}{1/2}$ \\
$\eta_s$ & $-1$ & $-1$ & $+1$ & $-1$ \\
\bottomrule
\end{tabular}
\end{center}
\vspace{-4mm}
\caption{Spin structures can be expressed in $(\alpha,\beta)$ or $s$.}
\label{table:spinstructures}
\end{table}
\begin{table}[t]
\renewcommand{\arraystretch}{1.2}
 \[
\begin{array}{ccccccccc}
\toprule
\sigma &   {\rm CP} &
\tilde{\chi} & \gamma_i & f(\gamma_i) & h_1 & h_2 & g_1 & g_2 \\
\midrule
\ck_u &  1 &
1 &2 k v_i & -2 \sin(\pi \gamma_i) 
& 0 &  0  & 0 & 0 \\
\ck_t & 1 &
 \tilde{\chi}(\Theta^{N/2},\Theta^k)&2 k v_i &1 
& \frac{1}{2} &  -\frac{1}{2}  & 0 & 0 \\
\ca_{99} & ({\rm tr} \gamma_9^k)^2 &
1 &  k v_i &-2 \sin(\pi \gamma_i) & 0 &  0  & 0 & 0 \\
\ca_{55} & ({\rm tr} \gamma_5^k)^2& 
1 &  k v_i & -2 \sin(\pi \gamma_i) &  0 &  0  & 0 & 0 \\
\ca_{95} & ({\rm tr} \gamma_9^k)({\rm tr} \gamma_5^k)& 
2 &  k v_i &1 &  \frac{1}{2} &  -\frac{1}{2}  & 0 & 0 \\
\cm_{9} & {\rm tr} \gamma_9^{2 k} &
- 1  & k v_i &-2 \sin(\pi \gamma_i) & 0 &0 &0 & 0 \\ 
 \cm_{5} & {\rm tr} \gamma_5^{2 k} & 
 -1 &k v_i & 2 \cos(\pi \gamma_i)& 0 & 0 &\frac{1}{2}& -\frac{1}{2} \\
\bottomrule
\end{array}
\]
\caption{Constants associated with partition functions. $\K_u$ and $\K_t$ denote the Klein bottle contributions with untwisted ($\K_u$) and $\Theta^{N/2}$-twisted ($\K_t$) closed strings running in the loop. $\tilde{\chi}(\Theta^{N/2},\Theta^k)$ for $\K_t$ denotes the number of simultaneous  fixed points of $\Theta^{N/2}$ and $\Theta^k$, see (A.4) of \cite{Font:1989aj}. The CP factors corresponding to the D5-branes assume that all D5-branes are sitting at the fixed point at the origin of the compact transverse space.}
\label{tab:details}
\end{table}

The expression \eqref{Z_sigma} is strictly speaking only valid for $\N=1$ sectors. For $\N\geq$2 sectors, one can use the following prescription \cite{Aldazabal:1998mr}. These sectors have the feature that at least along one torus $h_i$ vanishes and $\gamma_i+g_i$ is integer. In that case, \eqref{Z_sigma} has a well defined limit but one also has to include a sum over momentum or winding states. Concretely, one should perform the following substitutions:
\begin{itemize}
\item $\M_9$
\be
\gamma_i = \textrm{integer},\, i=1,2,3: \quad \frac{-2\sin \pi \gamma_i}{\vt\ss{\frac{1}{2}}{\frac{1}{2}+\gamma_i}} \rightarrow 
\frac{1}{\eta^3}\,\mathcal{L}^{[i,{\rm M}]} \label{m9subs}
\ee
\item $\M_5$
\be
&\gamma_i = \textrm{half-integer},\, i=1,2 &: \quad \frac{2\cos \pi \gamma_i}{\vt\ss{\frac{1}{2}}{\frac{1}{2}+\gamma_i+g_i}} = (-1)^i \frac{-2\sin\pi (\gamma_i+g_i)}{\vt\ss{\frac{1}{2}}{\frac{1}{2}+\gamma_i+g_i}} \rightarrow 
\frac{(-1)^i}{\eta^3}\,\mathcal{L}^{[i,{\rm W}]} \\
&\gamma_3 = \textrm{integer} &: \quad \frac{-2\sin \pi \gamma_3}{\vt\ss{\frac{1}{2}}{\frac{1}{2}+\gamma_3}} \rightarrow 
\frac{1}{\eta^3}\,\mathcal{L}^{[3,{\rm M}]}
\ee
\item $\A_{99}$
\be
\gamma_i = \textrm{integer},\, i=1,2,3: \quad \frac{-2\sin \pi \gamma_i}{\vt\ss{\frac{1}{2}}{\frac{1}{2}+\gamma_i}} \rightarrow 
\frac{1}{\eta^3}\,\mathcal{L}^{[i,{\rm M}]}
\ee
\item $\A_{55}$
\be
&\gamma_i = \textrm{integer},\, i=1,2 &: \quad \frac{-2\sin \pi \gamma_i}{\vt\ss{\frac{1}{2}}{\frac{1}{2}+\gamma_i}} \rightarrow 
\frac{1}{\eta^3}\,\mathcal{L}^{[i,{\rm W}]} \\
&\gamma_3 = \textrm{integer} &: \quad \frac{-2\sin \pi \gamma_3}{\vt\ss{\frac{1}{2}}{\frac{1}{2}+\gamma_3}} \rightarrow 
\frac{1}{\eta^3}\,\mathcal{L}^{[3,{\rm M}]}
\ee
\item $\A_{95}$
\be
\gamma_3 = \textrm{integer} : \quad \frac{-2\sin \pi \gamma_3}{\vt\ss{\frac{1}{2}}{\frac{1}{2}+\gamma_3}} \rightarrow 
\frac{1}{\eta^3}\,\mathcal{L}^{[3,{\rm M}]}
\ee
\item $\K_u$  $(\gamma_i =2 k v_i)$
\be
& \gamma_i = \textrm{even-integer},\, i=1,2,3 &: \quad \frac{-2\sin \pi \gamma_i}{\vt\ss{\frac{1}{2}}{\frac{1}{2}+\gamma_i}} \rightarrow 
\frac{1}{\eta^3}\,\mathcal{L}^{[i,{\rm M}]}\\
& \gamma_i = \textrm{odd-integer},\, i=1,2,3 &: \quad \frac{-2\sin \pi \gamma_i}{\vt\ss{\frac{1}{2}}{\frac{1}{2}+\gamma_i}} \rightarrow 
\frac{1}{\eta^3}\,\mathcal{L}^{[i,{\rm W}]}
\ee
\item $\K_t$ $(\gamma_3 =2 k v_3)$
\be
& \gamma_3 = \textrm{even-integer}&: \quad \frac{-2\sin \pi \gamma_3}{\vt\ss{\frac{1}{2}}{\frac{1}{2}+\gamma_3}} \rightarrow 
\frac{1}{\eta^3}\,\mathcal{L}^{[3,{\rm M}]}\\
& \gamma_3 = \textrm{odd-integer} &: \quad \frac{-2\sin \pi \gamma_3}{\vt\ss{\frac{1}{2}}{\frac{1}{2}+\gamma_3}} \rightarrow 
\frac{1}{\eta^3}\,\mathcal{L}^{[3,{\rm W}]}\ . \label{ktsubs}
\ee
\end{itemize}
Here $ \mathcal{L}^{[j,{\rm M/W}]}$ is the momentum/winding sum along the $j$th torus (with volume $V_j$ and metric $g_{ab}^{[j]}$, cf.\ \eqref{torusmetric}) given by
\be
\mathcal{L}^{[j,{\rm M}]} &=& \frac{V_j}{4 \pi^2 \alpha' t} \sum_{m^1,m^2} e^{- \frac{\pi}{t} m^a m^b g_{ab}^{[j]}}\ ,  \label{L_M}\\
\mathcal{L}^{[j,{\rm W}]} &=& \frac{4 \pi^2 \alpha' } {V_j \,t} \sum_{w_1,w_2} e^{- \frac{\pi}{ t} w_a w_b g^{[j] ab}}\ . \label{L_W} 
\ee
For $\A$ and $\M$ the momentum sum $\mathcal{L}^{[j,{\rm M}]}$ appears if the $j$th torus is parallel to the branes whereas the winding sum $\mathcal{L}^{[j,{\rm W}]}$ appears if the $j$th torus is transversal to the branes. For $\K$ the situation is as follows: If $\gamma_j$ is even, the corresponding torus is not reflected. The orientation reversal $\Omega$, however, reverses the winding modes. Thus, only the momentum modes survive. On the other hand, if $\gamma_j$ is odd, the corresponding torus is reflected (i.e.\ $kv_j$ is half-integer). Combined with the orientation reversal $\Omega$, this leaves the winding modes along this torus invariant, cf.\ sec.\ 9.14.2 in \cite{Kiritsis:2007zza}. Note that the terms ``momentum'' and ``winding'' as used here refer to the open string channel. When writing down \eqref{L_M} and \eqref{L_W} we  performed a Poisson resummation, cf.\ \eqref{Poisson}, thus expressing the sums in the closed string channel.

In order to see these substitution rules in action, we give the explicit form of the partition function for the $\mathbb{Z}_6'$ orientifold in app.\ \ref{partfun}.\footnote{In order to make contact between \eqref{Z_sigma} and the formulas in app.\  \ref{partfun} you have to use $\mathrm{tr} \big((\gamma_{\Omega_k,9})^\mathrm{T}(\gamma_{\Omega_k,9})^{-1}\big)={\rm tr} \gamma_9^{2 k}$ and $\mathrm{tr} \big((\gamma_{\Omega_k,5})^\mathrm{T}(\gamma_{\Omega_k,5})^{-1}\big)=-{\rm tr} \gamma_5^{2 k}$, cf.\  eqs.\ (2.36) and (2.41) in \cite{Aldazabal:1998mr}. Moreover, at some places (in particular for $\cm_5$) one has to use $\vt\ss{\frac{1}{2}}{1+a}=-\vt\ss{\frac{1}{2}}{a}$, cf.\ eq.\ \eqref{shift}.}

The substitutions \eqref{m9subs} - \eqref{ktsubs} can be done after performing the spin-structure summation. Thus, the sum over spin structures in \eqref{1pt} can be performed using \eqref{Z_sigma} and \eqref{Z_s^vt} for the partition function. Then we need the formula (cf.\ eq.\ (130) in \cite{Berg:2004ek})
\be \label{spin-sum}
\sum_{s=\text{even}} 
\,Z_s^\vt\, \frac{\vt_s''(0)}{\vt_s(0)}=\sum_{i=1}^3 
\frac{\vt'\ss{1/2+h_i}{1/2+\gamma_i+g_i}(0)}{\vt\ss{1/2+h_i}{1/2+\gamma_i+g_i}(0)}\ .
\ee
With this, \eqref{1pt} reads
\be
(\delta E)_\sigma &=& -\frac{\pi (\alpha')^2}{32 N (4 \pi^2 \alpha')^2}\,\int_0^\infty \frac{d t}{t^2}\, \frac{{\rm Im}(\tau_\sigma)}{t}
 \sum_{k=0}^{N-1}{\rm CP}_\sigma \tilde{\chi}_\sigma  \sin(\pi \gamma_3) \left(\prod_{j=1}^2 f(\gamma_j) \right) \sum_{i=1}^3 
\frac{\vt'\ss{1/2+h_i}{1/2+\gamma_i+g_i}(0)}{\vt\ss{1/2+h_i}{1/2+\gamma_i+g_i}(0)}\ .\nonumber \\ \label{delta-after-spin_sum}
\ee
Again this expression is strictly valid only for $\N=1$ sectors. $\N=4$ sectors vanish and for $\N=2$ sectors one would have to perform the substitutions mentioned above. In these sectors there is no contribution from massive string states, there are no theta functions in the integrand and, thus, the $t$-integral is relatively simple. We will see this in a concrete example in sec.\ \ref{z6n2}. 

Let us now look at the contributions from $\N=1, 2$ sectors in turn and let us see how far we can get without specializing to a concrete model. 


\subsection{$\N=1$ sectors}


The $\N=1$ sectors can be treated in a way very analogous to secs.\ 3.8 - 3.11 of \cite{Berg:2014ama}. Their contribution to the Planck mass is given by 
\be
(\delta E)^{({\cal N} = 1)} = \sum_\sigma (\delta E)_\sigma^{({\cal N} = 1)} = - \frac{\pi (\alpha')^2}{64 N (4 \pi^2 \alpha')^2}\,\int_0^\infty \frac{d t}{t^2}\,  \sum_\sigma \sum_{k\in \{\N=1 \}} {\rm CP}_\sigma\,\sigma^{(k)}. \label{delta-sigma}
\ee
Here\footnote{Note that $\sigma$ in summations and subscripts stands for a surface, i.e.\ $\sigma=\{\ck_u, \ck_t, \ca_{95},\ca_{99},\ca_{55},\cm_9,\cm_5\}$, but when we write $\sigma$ with superscript $^{(k)}$ such as in $\cm_5^{(k)}$, we mean quantities defined in \eqref{sigma^k}.}
\be
\sigma^{(k)}= \tilde{e}_\sigma\, \tilde{\chi}_\sigma  \sin(\pi \gamma_3) \left(\prod_{j=1}^2 f(\gamma_j) \right)  \hat{\sigma}^{(k)}
\quad \textrm{for $k \in \{\N=1 \}$} 
\label{sigma^k}
\ee
with
\be
 \tilde{e}_\sigma = 
 \begin{cases}
 1 & \text{for } \A,\M \\
 4 & \text{for } \K
 \end{cases} \,,
\ee
and 
\be
\hat{\sigma}^{(k)}= \sum_{i=1}^3 
\frac{\vt'\ss{1/2+h_i}{1/2+\gamma_i+g_i}(0)}{\vt\ss{1/2+h_i}{1/2+\gamma_i+g_i}(0)}\ . \label{hat-sigma}
\ee
For later use, we also introduce 
\be
 e_\sigma = 
 \begin{cases}
 1 & \text{for } \A \\
 4 & \text{for } \M,\K
 \end{cases} \, , \label{e_sigma}
\ee
which differs from $\tilde{e}_\sigma$ for the M\"obius amplitude.

From \eqref{sigma^k}-\eqref{hat-sigma} and table \ref{tab:details}, we have
\be
\ck_u^{(k)} &=&  16\, \sin(2 \pi\,k\,v_3) \sin(2 \pi\,k\,v_1) \sin(2 \pi\,k\,v_2) \,\hat{\ck}_u^{(k)} \ , \label{ck_u^k}\\
\ca_{99}^{(k)} &=& 4\, \sin(\pi\,k\,v_3) \sin(\pi\,k\,v_1) \sin(\pi\,k\,v_2) \,\hat{\ca}_{99}^{(k)}\ , \\
\cm_9^{(k)}&=& -4\, \sin(\pi\,k\,v_3) \sin(\pi\,k\,v_1) \sin(\pi\,k\,v_2) \,\hat{\cm}_9^{(k)}\ , \label{cm^k}\\
\ck_{t}^{(k)}&=& 4 \, \tilde{\chi}(\Theta^{N/2},\Theta^k)\, \sin(2 \pi\,k\,v_3)\, \hat{\ck}_{t}^{(k)} \label{ck^k}\ , \\
\ca_{55}^{(k)} &=& 4\, \sin(\pi\,k\,v_3) \sin(\pi\,k\,v_1) \sin(\pi\,k\,v_2) \,\hat{\ca}_{55}^{(k)}\ , \\
\ca_{95}^{(k)}&=&2\, \sin(\pi\,k\,v_3) \,\hat{\ca}_{95}^{(k)}\ , \label{ca^k} \\
\cm_5^{(k)}&=& -4 \,\sin(\pi\,k\,v_3) \cos(\pi\,k\,v_1) \cos(\pi\,k\,v_2)\, \hat{\cm}_5^{(k)}\ .  \label{cm5^k}
\ee 
Note that for odd $N$ there is no contribution from $\ck_{t}$, $\ca_{55}$, $\ca_{95}$ and $\cm_5$.

Using the behavior of the theta functions under shifts in their characteristics, cf.\ eq.\ \eqref{shift}, and the fact that the even/odd spin structure theta functions are even/odd functions of their argument, together with the supersymmetry condition $\sum_i v_i=0$, one can check that
\be
 \hat{\sigma}^{(q N\pm k)}=\pm \hat{\sigma}^{(k)} \quad \text{for all $\sigma$}, &&\quad 
\hat{\sigma}^{\left(\frac{q N}{2} \pm k\right)}=\pm \hat{\sigma}^{(k)} \quad \text{for}\ \K\ , \\
 \sigma^{(q N\pm k)}= \sigma^{(k)} \quad \text{for all $\sigma$}, &&\quad 
\sigma^{\left(\frac{q N}{2} \pm k\right)}= \sigma^{(k)} \quad \text{for}\ \K\ .  \label{property-sigma^k}
\ee
Here $q$ is an arbitrary integer and $N$ is the order of the orbifold group $\mathbb{Z}_N$. These identities allow the individual sectors to be related to each other. We will make use of this in the examples below, cf.\ secs.\ \ref{z3} and \ref{z6}.

For $\N=1$ sectors with $h_i = 0$ the $t$-integral in eq.\ \eqref{delta-sigma} can be performed using the results of \cite{Berg:2014ama} (cf.\ (115)-(117)), i.e.\ (assuming $0<\gamma<1$ for $\A$ and $\K$, and $0<\gamma<1/2$ for $\M$)\footnote{The surface dependent cutoffs at the lower end of the $t$-integrals ensure a uniform cutoff in the closed string channel, i.e.\ $\ell = 1/(e_\sigma t) < \Lambda$.}
\begin{align}
I_{\ca/\ck}(\gamma)= \int_{\frac{1}{e_\sigma \Lambda}}^\infty \frac{\dd t}{t^2} \,
 \frac{\vartheta_1'(\gamma,\tau_\sigma )}{\vartheta_1(\gamma, \tau_\sigma)}
 & = e_\sigma \pi(1-2\gamma) \Lambda^2 + e_\sigma \frac{\pi}{24} \left[ \psi'(\gamma) - \psi'(1-\gamma) \right] \,, \label{I_A_K} \\
I_\cm(\gamma)= \int_{\frac{1}{4 \Lambda}}^\infty \frac{\dd t}{t^2} \,
 \frac{\vartheta_1'(\gamma,\hf+\frac{it}{2})}{\vartheta_1(\gamma,\hf+\frac{it}{2})}
 & = 8\pi(1-4\gamma) \Lambda^2 + \frac{\pi}{12} \left[ \psi'(\gamma) - \psi'(1-\gamma)- \thf \psi'(\thf+\gamma) + \thf \psi'(\thf-\gamma) \right] \,.
 \label{I_M}
\end{align}
Here $\psi'(x)$ denotes the trigamma function, i.e.\ the derivative of the digamma function $\psi(x) = {\Gamma'(x)}/{\Gamma(x)}$.

The $t$-integral of terms with $h_i =\pm 1/2$, appearing in $\K_t$ and $\ca_{95}$, is new. We compute it in app.\ \ref{t-integral} where we find (again for $0<\gamma<1$)
\be
\tilde{I}_{\ca/\ck}(\gamma)=\int_{\frac{1}{e_\sigma \Lambda}}^\infty \frac{\dd t}{t^2} \,
 \frac{\vartheta_4'(\gamma,\tau_\sigma )}{\vartheta_4(\gamma, \tau_\sigma)}
  = e_\sigma \pi(1-2\gamma) \Lambda^2 - e_\sigma \frac{\pi}{48} \left[ \psi'(\gamma) - \psi'(1-\gamma) \right] \,. \label{tildeI_A_K}
\ee 


\subsection{$\N=2$ sectors}

$\N=2$ sectors are characterized by the fact that along exactly one torus (say the $n$th torus) $h_n$ vanishes and $\gamma_n+g_n$ is integer. In this case, one has to take a limit of \eqref{delta-after-spin_sum}. It is clear from table \ref{tab:details} that $\sin(\pi \gamma_3) \left(\prod_{j=1}^2 f(\gamma_j) \right)$ vanishes in this case and the only contribution to $(\delta E)_\sigma$ comes from the summand with $i=n$, i.e.\ from the term with $\vt\ss{1/2+h_n}{1/2+\gamma_n+g_n}(0)$ in the denominator which also vanishes.\footnote{On the other hand, for $\N=4$ sectors $h_i$ vanish and $\gamma_i+g_i$ are integer along all three tori. Thus, the numerator of \eqref{delta-after-spin_sum} has a triple zero which can not be balanced by the simple zero in the denominator. Consequently the $\N=4$ sectors do not contribute. \label{fn_neq4}} Then the substitution rules \eqref{m9subs} - \eqref{ktsubs} lead to 
\be
\big(-2 \sin\pi(\gamma_n+g_n)\big) \, \frac{\vt'\ss{1/2}{1/2+\gamma_n+g_n}(0)}{\vt\ss{1/2}{1/2+\gamma_n+g_n}(0)}  &\to& 
\frac{\vt'\ss{\frac{1}{2}}{\frac{1}{2}+\gamma_n+g_n}(0)}{\eta^3} \mathcal{L}^{[n,{\rm M/W}]} \\
\nonumber \\
& = & (-2 \pi) (-1)^{\gamma_n+g_n} \mathcal{L}^{[n,{\rm M/W}]}\ .
\ee

To summarize, the $\N=2$ sector contribution is given by
\be
(\delta E)^{(\N=2)}=\sum_{\sigma} (\delta E)_\sigma^{(\N=2)} = - \frac{\pi (\alpha')^2}{64 N (4 \pi^2 \alpha')^2}\, \int_0^\infty \frac{d t}{t^2}\,  
\sum_{\sigma} \sum_{k\in \{\N=2 \}} {\rm CP}_\sigma\,\sigma^{(k)}\ . \label{delta-2}
\ee
Here
\be
\sigma^{(k)}=  \pi\,  \tilde{e}_\sigma\, \tilde{\chi}_\sigma  D_\sigma^{(k)} \, \mathcal{L}^{[n,{\rm M/W}]} \quad \textrm{for $k \in \{\N=2 \}$} 
\label{sigma^k-2} 
\ee
with the constant factor $D_\sigma^{(k)}$ given by 
\be
D_\sigma^{(k)} = (-1)^{\gamma_n+g_n} \prod_{i \neq n}^3 F(\gamma_i) \label{D}
\ee
with
\be
F(\gamma_i) = \begin{cases}
  f(\gamma_i)     & \quad \text{for $i=1$ and $2$} \\
 -2 \sin \pi \gamma_3    & \quad \text{for $i=3$}  \\
  \end{cases}\ . \label{Fgammai}
\ee
Obviously, $n$ depends on the concrete $\N = 2$ sector, i.e.\ on $k$ and $\sigma$. 

Let us express \eqref{L_M} and \eqref{L_W} collectively as
\be
\mathcal{L}^{[n,{\rm M/W}]} = \frac{C^{[n,{\rm M/W}]}}{t} \sum_{m^1, m^2} 
e^{- \frac{\pi}{t} m^a m^b\, g_{ab}^{[n,{\rm M/W}]}}\ ,  \label{L_M_W}
\ee
where
\be
C^{[n,{\rm M/W}]}  =  \begin{cases}
  \frac{V_n}{4 \pi^2 \alpha'}       & \quad \text{for M (momentum sum)} \\
  \frac{4 \pi^2 \alpha'}{V_n}    & \quad \text{for W (winding sum)}\\
  \end{cases} \label{C_M-W}
\ee
and
\be
g_{ab}^{[n,{\rm M/W}]}= \begin{cases}
g_{ab}^{[n]}      & \quad \text{for M (momentum sum)} \\
g^{[n] ab}     & \quad \text{for W (winding sum)}\\
    \end{cases} \ , \label{g_M-W}
\ee
i.e.\ $g_{ab}^{[n, {\rm W}]}$ is the inverse matrix of $g_{ab}^{[n, {\rm M}]}$.

Now we split $\mathcal{L}^{[n,{\rm M/W}]}$ as
\be
\mathcal{L}^{[n,{\rm M/W}]} &=& \frac{C^{[n,{\rm M/W}]}}{t} \left(1+ 
\sum_{\vec{m} \in \mathbb{Z}^2 \setminus \vec{0}} 
e^{- \frac{\pi}{t} m^a m^b\, g_{ab}^{[n,{\rm M/W}]}}\right)  \\
&=& \frac{C^{[n,{\rm M/W}]}}{t} + \mathcal{L}'^{[n,{\rm M/W}]}
\ee
with 
\be
\mathcal{L}'^{[n,{\rm M/W}]} = \frac{C^{[n,{\rm M/W}]}}{t}  \sum_{\vec{m} \in \mathbb{Z}^2 \setminus \vec{0}}  
e^{- \frac{\pi}{t} m^a m^b\, g_{ab}^{[n,{\rm M/W}]}}\ .
\ee
Then we have
\be
\int_{\frac{1}{e_\sigma \Lambda}}^{\infty} \frac{\dd t}{t^2} \mathcal{L}^{[n,{\rm M/W}]}
=  \frac{C^{[n,{\rm M/W}]} \,e_\sigma^2 \Lambda^2}{2} + \int_{0}^{\infty} \frac{\dd t}{t^2} \mathcal{L}'^{[n,{\rm M/W}]}\ .
\label{t-integral-L}
\ee
Here we set $\Lambda=\infty$ in the second term on the right hand side since it is finite in the limit $\Lambda \rightarrow \infty$. It can be evaluated using (see app.\ \ref{t-integral_n2})
\bea
\Gamma^{[n,{\rm M/W}]} & \equiv & \int_{0}^{\infty} \frac{\dd t}{t^3}  \sum_{\vec{m} \in \mathbb{Z}^2 \setminus \vec{0}}  
e^{- \frac{\pi}{t} m^a m^b\, g_{ab}^{[n,{\rm M/W}]}} \label{Gammadef} \\ 
& = & \begin{cases}
\frac{(4 \pi^2 \alpha')^2 }{\pi^2 V_n^2}\,  E_2\big(U^{[n]}\big),  & \quad \text{for M (momentum sum)}\\ \\
\frac{V_n^2}{\pi^2 ( 4 \pi^2 \alpha')^2 }\, E_2\left(- \frac{1}{U^{[n]}}\right) & \quad \text{for W (winding sum)} \\ 
\end{cases}\ , \label{Gamma_E}
\eea
where $U^{[n]}$ is the complex structure of the $n$th torus and $E_2$ is a non-holomorphic Eisenstein series, cf.\ \eqref{eisenstein}.

Now we collected all the relevant formulas to evaluate the 1-loop correction to the Planck mass in explicit models. For illustration we do so for one odd and one even order orbifold group, i.e.\ for the $\mathbb{Z}_3$ and $\mathbb{Z}_6'$ orientifolds.


\section{Example: $\mathbb{Z}_3$}
\label{z3}

Let us begin with the $\mathbb{Z}_3$ orientifold, which is the simplest example of $\mathbb{Z}_N$ with odd $N$. 
This has the twist vector $ v=\left(\frac{1}{3},\frac{1}{3},-\frac{2}{3}\right)$ and only D9-branes (no D5-branes). Furthermore, there are no $\N=2$ sectors. As discussed in footnote \ref{fn_neq4} above, the $\N=4$ sector (i.e.\ $k=0$) vanishes, so we are left with only $\N=1$ sector contributions. 

Their contribution to the Planck mass is determined by (cf.\ \eqref{delta-sigma})
\be
\sum_{\sigma} \sum_{k \in \{\N=1 \}} {\rm CP}_\sigma \, \sigma^{(k)} &=& \sum_{k=1,2} \left[ 
\ck_u^{(k)} + (\mbox{tr} \gamma_9^k)^2 \ca_{99}^{(k)}+ (\mbox{tr} \gamma_9^{2k}) \cm_9^{(k)} \right]\\
&=& 2\left[ 
\ck_u^{(1)} + 16 \ca_{99}^{(1)} -4 \cm_9^{(1)} \right]  \\
&=& 32 \left(\prod_{j=1}^3 \sin \pi v_j\right) \left[ 
- \hat{\ck}_u^{(1)} + 4 \hat{\ca}_{99}^{(1)} + \hat{\cm}_9^{(1)} \right]\ .  \label{k-sum-3} 
\ee
In the second equality we used \eqref{property-sigma^k} and the tadpole conditions $|\mbox{tr} \gamma_9| = 4$ and $\mbox{tr} \gamma_9^2 = \mbox{tr} \gamma_9^4  = -4$ (cf.\ (2.37) and the line below that eq.\ in \cite{Aldazabal:1998mr}).
In the third equality we used \eqref{ck_u^k}-\eqref{cm^k}.

Next we have to perform the $t$-integral, i.e.
\be
\int_0^\infty \frac{\dd t}{t^2}\left[ - \hat{\ck}_u^{(1)} + 4 \hat{\ca}_{99}^{(1)} + \hat{\cm}_9^{(1)} \right] &=& 
- \int_{\frac{1}{4 \Lambda}}^\infty \frac{\dd t}{t^2}  \hat{\ck}_u^{(1)} + 4 \int_{\frac{1}{\Lambda}}^\infty \frac{\dd t}{t^2}  \hat{\ca}_{99}^{(1)} 
+ \int_{\frac{1}{4 \Lambda}}^\infty \frac{\dd t}{t^2}  \hat{\cm}_9^{(1)}\, \nonumber \\ 
&=& 
- 3 \int_{\frac{1}{4 \Lambda}}^\infty \frac{\dd t}{t^2} \frac{\vt'_1(2 v_1,\tau_\ck)}{\vt_1(2 v_1,\tau_\ck)}
+ 12 \int_{\frac{1}{\Lambda}}^\infty \frac{\dd t}{t^2}  \frac{\vt'_1(v_1,\tau_\ca)}{\vt_1( v_1,\tau_\ca)}+ \nonumber \\ 
&&  + 3 \int_{\frac{1}{4 \Lambda}}^\infty \frac{\dd t}{t^2}  \frac{\vt'_1(v_1,\tau_\cm)}{\vt_1( v_1,\tau_\cm)} \nonumber \\ 
\nonumber \\
&=& -3 I_\ck(2 v_1) + 12 I_\ca(v_1) + 3 I_\cm(v_1)  \nonumber \\ \nonumber \\
&=& \Lambda^2 \left(\frac{12 \pi}{3} + \frac{12 \pi}{3} -\frac{24 \pi}{3} \right) + \nonumber \\
&& +  \frac{5 \pi}{4}\Big[\psi'\left(\frac{1}{3}\right)-\psi'\left(\frac{2}{3}\right)\Big]
+ \frac{\pi}{8}\Big[\psi'\left(\frac{1}{6}\right)-\psi'\left(\frac{5}{6}\right)\Big] 
  \nonumber \\
   \nonumber \\
&=& \frac{15 \pi}{4} \Big[\psi'\left(\frac{1}{3}\right)- \frac{2 \pi^2}{3} \Big]  \nonumber \\
&=& 15 \pi\, \sin\left(\frac{\pi}{3}\right) {\rm Cl}_2\left(\frac{\pi}{3}\right). \label{finite-3}
\ee
In the second equality we used that for $\mathbb{Z}_3$ (i.e.\ for $ v=\left(\frac{1}{3},\frac{1}{3},-\frac{2}{3}\right)$) the quantities $\hat{\sigma}^{(1)}$ of eq.\ \eqref{hat-sigma} (with $\sigma=\{\ck_u, \ca_{99},\cm_9\}$) can be simplified to
\be
\hat{\sigma}^{(1)} &=& 2 \frac{\vt'\ss{1/2}{1/2+\gamma_1}}{\vt\ss{1/2}{1/2+\gamma_1}} + \frac{\vt'\ss{1/2}{1/2+\gamma_3}}{\vt\ss{1/2}{1/2+\gamma_3}} \\
&=& 3 \frac{\vt'\ss{1/2}{1/2+\gamma_1}}{\vt\ss{1/2}{1/2+\gamma_1}}\ .
\ee 
In the fourth equality we used \eqref{I_A_K} and \eqref{I_M}, while the fifth and sixth equalities make use of 
\be
\psi'\left(\frac{2}{3}\right) = - \psi'\left(\frac{1}{3}\right) + \frac{4\pi^2}{3} \quad , \quad 
\psi'\left(\frac{1}{6}\right) =  5 \psi'\left(\frac{1}{3}\right) -\frac{4\pi^2}{3} \quad , \quad 
\psi'\left(\frac{5}{6}\right) = - 5\psi'\left(\frac{1}{3}\right) + \frac{16\pi^2}{3} \label{psiprime}
\ee
and
\be
\psi'\left(\frac{1}{3}\right) = 4\sin \left(\frac{\pi}{3}\right) \Cl_2\left(\frac{\pi}{3}\right) + \frac{2\pi^2}{3}\ , \label{psicl2}
\ee
respectively. Here ${\rm Cl}_2$ is the second Clausen function. Note that the UV divergences ($\propto \Lambda^2$) cancel.  

Putting all constant factors together, the final result reads (using \eqref{delta-sigma}, \eqref{k-sum-3} and \eqref{finite-3})
\be
(\delta E)_{\K+\A+\M} &=& 
- \frac{ \pi (\alpha')^2}{64 N (4 \pi^2 \alpha')^2}\,32 \sin \left(\frac{\pi}{3} \right)^3 \sin \left(-\frac{2\pi}{3} \right) \,15 \pi \, {\rm Cl}_2\left(\frac{\pi}{3}\right) \nonumber \\
&=& \frac{45}{512\, \pi^2}\, {\rm Cl}_2\left(\frac{\pi}{3}\right) \label{delta_Z3}\ .
\ee
To this one still has to add the contribution from the sphere and the torus, cf.\ \eqref{deltaETS}, leading to 
\be
\delta E = \frac{45}{512\, \pi^2}\, {\rm Cl}_2\left(\frac{\pi}{3}\right) + \frac{\chi}{(2 \pi)^3} \Big( 2 \zeta (3) \frac{e^{-2 \Phi_4}}{{\cal V}} + \frac{\pi^2}{3} \Big)\label{delta_Z3_final}\ ,
\ee
where ${\rm Cl}_2 (\pi/3) \approx 1.015$ and the Euler number of the $\mathbb{Z}_3$ orientifold is $\chi = 2 (h^{(1,1)}-h^{(2,1)}) = 72$, cf.\ table 20 in \cite{Blumenhagen:2006ci}. 

There is a relation between the Clausen function and the Hurwitz zeta function, i.e.
\be
\zeta(2, 5/6)-\zeta(2,1/6) &=& \psi'(5/6) - \psi'(1/6) \\
&=& -10 \left[\psi'(1/3) -\frac{2 \pi^2}{3}\right] \\
&=& - 40 \sin(\pi/3) {\rm Cl}_2\left(\frac{\pi}{3}\right) \\
&=& - 20 \sqrt{3}\, {\rm Cl}_2\left(\frac{\pi}{3}\right)\ .
\ee
Here we used $\zeta(1+n, \gamma) = \frac{(-1)^{n+1}}{n!} \psi^{(n)}(\gamma)$ in the first line, and \eqref{psiprime} and \eqref{psicl2} in the second and third lines, respectively. This relation shows that our result \eqref{delta_Z3} is very similar to the type IIA result found by Epple, cf.\ (3.10) in \cite{Epple:2004ra}. The overall coefficients do not match, but this is not too surprising, given that the IIA and IIB $\mathbb{Z}_3$ orientifolds are not T-dual to each other.\footnote{Note, however, footnote \ref{signdiff}; moreover, there is an overall factor of $\pi$ missing on the right hand side of formula (A.15) in \cite{Epple:2004ra} and the sign in the middle expression of (3.10) in  \cite{Epple:2004ra} is wrong, as can be seen from eq.\ (B.8) therein. This is also clear from the fact that $\zeta(2,5/6) -\zeta(2,1/6) \approx -35.16$ is negative, wheras the right hand side of (3.10) is positive.} Rather, as we already mentioned in the introduction, under T-duality the type IIB orientifold we are discussing here would be mapped to an {\it asymmetric} type IIA orientifold.

Note also that our result differs from the one found in \cite{Kohlprath:2003pu} for non-compact type IIB orientifolds with odd $N$ and only D3-branes (instead of D9-branes in our case). There the conclusion was that the overall contribution of $\A, \M$ and $\K$ vanishes, due to a cancellation between the $k$ and $(N-k)$ sectors. This discrepancy can be traced back to the fact that \cite{Kohlprath:2003pu} uses absolute values of the sin-factors in the partition function. Our understanding is that the absolute values should only appear in the $t \rightarrow 0$ limit, cf.\ (7.12) and (7.14) in \cite{Aldazabal:1998mr}, for instance.



\section{Example: $\mathbb{Z}_6'$}
\label{z6}

The $\mathbb{Z}_6'$ orientifold has twist vector $v=\left(\frac{1}{6},-\frac{1}{2},\frac{1}{3}\right)$. Given that the torus lattice has to be invariant under the orbifold action, the complex structures of the first and third torus are fixed, whereas the complex structure of the second torus is still a free modulus $U_2$. The model has both D9-branes and D5-branes wrapped around the third torus. For simplicity we assume that all the D5-branes are sitting at the fixed point at the origin of the compact transverse space. Moreover, in addition to the $\N=1$ and $\N=4$ sectors which were already present in the last example of $\mathbb{Z}_3$, it also features $\N=2$ sectors. The different sectors are shown in table \ref{tab:Vol} which also indicates the volume dependence of the different sectors ($V_j$ stands for the volume of the $j$th torus). $\N=2$ sectors exhibit a single volume factor, $\N=4$ sectors three volume factors and $\N=1$ sectors only get contributions from completely localized strings so that they do not sense any of the torus volumes (i.e.\ they correspond to empty fields in table \ref{tab:Vol}). 
\begin{table}[h]
\renewcommand{\arraystretch}{1.2} 
\[
\begin{array}{|c||c|c|c|c|c|c|}
\hline
\sigma \setminus k & 0 & 1 & 2 & 3 & 4 & 5  \\
\hline
\hline
\ck_u & V_1 V_2 V_3 & \frac{1}{V_2} & V_2
& \frac{V_3}{V_1 V_2} &  V_2  & \frac{1}{V_2}  \\
\hline
\ck_t & V_3 &  &  & V_3 &    &   \\
\hline
\ca_{99} & V_1 V_2 V_3 &   & V_2 & V_3  & V_2&   \\
\hline
\ca_{55} &  \frac{V_3}{V_1 V_2}&   &  \frac{1}{V_2} &  V_3 &  \frac{1}{V_2}  &  \\
\hline
\ca_{95} & V_3 &  &  & V_3 &    &  \\
\hline
\cm_{9} &V_1 V_2 V_3 &  & V_2 & V_3 &V_2 &  \\ 
\hline
 \cm_{5} & V_3 & \frac{1}{V_2} &  & \frac{V_3}{V_1 V_2} &  &\frac{1}{V_2} \\
\hline
\end{array}
 \]
\caption{Volume factors for the different sectors of the $\mathbb{Z}_6'$ orientifold. Fields with no entry correspond to $\N = 1$ sectors, fields with a single volume factor correspond to $\N = 2$ sectors and fields with three volume factors denote $\N = 4$ sectors. Volumes in the numerator/denominator are accompanied by momentum/winding sums.} 
\label{tab:Vol}
\end{table}
%


\subsection{${\cal N}=1$ sectors}
\label{z6n1}
 
The $\N=1$ sector sum in \eqref{delta-sigma} for $\mathbb{Z}_6'$ is given by
\bea
&& \sum_{\sigma} \sum_{k \in \{\N=1 \}} {\rm CP}_\sigma \, \sigma^{(k)}=  \\
&&=\sum_{k=1,2,4,5} \ck_{t}^{(k)} + \sum_{k=2,4} \left[ (\mbox{tr} \gamma_9^k) (\mbox{tr} \gamma_5^k) \ca_{95}^{(k)} \right] + \sum_{k=1,5} (\mbox{tr} \gamma_9^{2k}) \cm_{9}^{(k)} + \sum_{k=2,4} (\mbox{tr} \gamma_5^{2k}) \cm_{5}^{(k)} \\
&&\,\, + \sum_{k=1,5}  \left[ (\mbox{tr} \gamma_9^k) (\mbox{tr} \gamma_5^k) \ca_{95}^{(k)} \right]  + \sum_{k=1,5} \left[(\mbox{tr} \gamma_9^k)^2 \ca_{99}^{(k)}  + (\mbox{tr} \gamma_5^k)^2 \ca_{55}^{(k)}\right]   \\
&&=\sum_{k=1,2,4,5} \ck_{t}^{(k)} + \sum_{k=2,4} \left[ (\mbox{tr} \gamma_9^k) (\mbox{tr} \gamma_5^k) \ca_{95}^{(k)} \right] + \sum_{k=1,5} (\mbox{tr} \gamma_9^{2k}) \cm_{9}^{(k)} + \sum_{k=2,4} (\mbox{tr} \gamma_5^{2k}) \cm_{5}^{(k)}. 
\eea
In the last equality we used the tadpole condition $\mbox{tr}(\gamma_9^k)=0=\mbox{tr}(\gamma_5^k)$ for $k=1, 5$.
Using \eqref{property-sigma^k}, the Chan-Paton traces $\mbox{tr} \gamma_{9}^2 =\mbox{tr} \gamma_{5}^2 =-8$, 
$\mbox{tr} \gamma_{9}^4 =\mbox{tr} \gamma_{5}^4 =8$ and $ \gamma_{9}^6 = \gamma_{5}^6 =-1$ \cite{Aldazabal:1998mr} and $\tilde{\chi}(\Theta^3,\Theta^k)=4$ for $k=1,2,4,5$, we obtain
\be
  \sum_{\sigma} \sum_{k \in \{\N=1 \}} {\rm CP}_\sigma\, \sigma^{(k)}&=&
4 \ck_t^{(1)} + 128 \ca_{95}^{(2)} - 16 \cm_9^{(1)} +16 \cm_5^{(2)} \\
&=& - 32 \sin( \pi v_3) \left[ -2 \hat{\ck}_t^{(1)} - 8 \hat{\ca}_{95}^{(2)} + \hat{\cm}_9^{(1)} - \hat{\cm}_5^{(2)}\right]  \\
&=& - 64 \sin( \pi v_3) \left[ - \hat{\ck}_t^{(1)} - 4 \hat{\ca}_{95}^{(2)} + \hat{\cm}_9^{(1)} \right]  \label{k-sum} \;. 
\ee
In the second and third equality we used \eqref{cm^k}-\eqref{cm5^k} and $\hat{\cm}_5^{(2)}=-\hat{\cm}_9^{(1)}$ (as can be shown  from \eqref{hat-sigma} and  \reftab{tab:details}), respectively.

Let us look at 
\be
\int_0^\infty \frac{\dd t}{t^2}\left[ - \hat{\ck}_t^{(1)} - 4 \hat{\ca}_{95}^{(2)} + \hat{\cm}_9^{(1)} \right] = 
- \int_{\frac{1}{4 \Lambda}}^\infty \frac{\dd t}{t^2}  \hat{\ck}_t^{(1)} - 4 \int_{\frac{1}{\Lambda}}^\infty \frac{\dd t}{t^2}  \hat{\ca}_{95}^{(2)} 
+ \int_{\frac{1}{4 \Lambda}}^\infty \frac{\dd t}{t^2}  \hat{\cm}_9^{(1)}. 
\ee
Then using \eqref{hat-sigma} and \eqref{I_A_K} - \eqref{tildeI_A_K}, we obtain\footnote{Note that the theta terms in  \eqref{hat-sigma} along the second torus (i.e.\ for $i=2$) vanish (even before integrating), so that we disregard the $i=2$ terms from now on.}
\be
&& - \int_{\frac{1}{4 \Lambda}}^\infty \frac{\dd t}{t^2}  \hat{\ck}_t^{(1)} - 4 \int_{\frac{1}{\Lambda}}^\infty \frac{\dd t}{t^2}  \hat{\ca}_{95}^{(2)} 
+ \int_{\frac{1}{4 \Lambda}}^\infty \frac{\dd t}{t^2}  \hat{\cm}_9^{(1)} \\
&& \quad = - \Big[\tilde{I}_{\ck}(2 v_1) + I_{\ck}(2 v_3) \Big]
-4 \Big[\tilde{I}_{\ca}(2 v_1) + I_{\ca}(2 v_3)\Big] + \Big[I_{\cm}(v_1)+I_{\cm}(v_3)\Big] \label{IplustildeI} \\
&& \quad = \sum_{i=1,3} \Big[- 4 \pi(1- 4 v_i) \Lambda^2 - 4 \pi(1- 4 v_i) \Lambda^2 + 8 \pi(1- 4 v_i) \Lambda^2  \Big] + \textrm{finite constant} \\
&& \quad = \textrm{finite constant}\ . 
\ee
Thus the UV divergences ($\propto \Lambda^2$) cancel and we are left with a finite constant. The finite contribution can be read off from \eqref{IplustildeI}, plugging in \eqref{I_A_K} - \eqref{tildeI_A_K}. This results in
\be
\frac{\pi}{8}\Big[\psi'\left(\frac{1}{6}\right)-\psi'\left(\frac{5}{6}\right)\Big] 
-\frac{5 \pi}{8}\Big[\psi'\left(\frac{2}{3}\right)-\psi'\left(\frac{1}{3}\right)\Big] = \frac{10 \pi}{4} \Big[\psi'\left(\frac{1}{3}\right)- \frac{2 \pi^2}{3} \Big] = 10 \pi \sin\left(\frac{\pi}{3}\right)\, {\rm Cl}_2\left(\frac{\pi}{3}\right)\ . \nonumber \\ \label{finite}
\ee
Putting all constant factors together, the final result reads (using \eqref{delta-sigma}, \eqref{k-sum} and \eqref{finite})
\be
(\delta E)^{(\N=1)} = \sum_\sigma (\delta E)_\sigma^{(\N=1)} &=& 
- \frac{\pi (\alpha')^2}{64 N (4 \pi^2 \alpha')^2}\,(-64) \sin \left(\frac{\pi}{3} \right)^2 \,10 \pi \, {\rm Cl}_2\left(\frac{\pi}{3}\right) \nonumber \\
&=& \frac{5}{64 \pi^2}\, {\rm Cl}_2\left(\frac{\pi}{3}\right)\ .
\ee


\subsection{${\cal N}=2$ sectors}
\label{z6n2}

Let us next consider the contribution from $\N=2$ sectors. Using \reftab{tab:details} and \reftab{tab:Vol}, it is given by
\bea
&& \sum_{\sigma} \sum_{k \in \{\N=2 \}} {\rm CP}_\sigma \, \sigma^{(k)}= \\
&&=\sum_{k=1,2,4,5} \ck_{u}^{(k)} +\sum_{k=0,3} \ck_{t}^{(k)} + \sum_{k=2,3,4} (\mbox{tr} \gamma_9^{2k}) \cm_{9}^{(k)} + \sum_{k=0,1,5} (\mbox{tr} \gamma_5^{2k}) \cm_{5}^{(k)} \nonumber \\
&&\,\,\, + \sum_{k=0,3}  \left[ (\mbox{tr} \gamma_9^k) (\mbox{tr} \gamma_5^k) \ca_{95}^{(k)} \right]  + \sum_{k=2,3,4} \left[(\mbox{tr} \gamma_9^k)^2 \ca_{99}^{(k)}  + (\mbox{tr} \gamma_5^k)^2 \ca_{55}^{(k)}\right]   \\
&&=\sum_{k=1,2,4,5} \ck_{u}^{(k)} +\sum_{k=0,3} \ck_{t}^{(k)} + \sum_{k=2,3,4} (\mbox{tr} \gamma_9^{2k}) \cm_{9}^{(k)} + \sum_{k=0,1,5} (\mbox{tr} \gamma_5^{2k}) \cm_{5}^{(k)} \nonumber \\
&&\,\,\, +   (\mbox{tr} \gamma_9^0) (\mbox{tr} \gamma_5^0) \ca_{95}^{(0)}   + \sum_{k=2,4} \left[(\mbox{tr} \gamma_9^k)^2 \ca_{99}^{(k)}  + (\mbox{tr} \gamma_5^k)^2 \ca_{55}^{(k)}\right] \label{2ndeq}\\
&&=  \sum_{k=0,3} \ck_t^{(k)} +(\mbox{tr} \gamma_9^0) (\mbox{tr} \gamma_5^0) \ca_{95}^{(0)} + (\mbox{tr} \gamma_9^{6}) \cm_9^{(3)}
+(\mbox{tr} \gamma_5^{0}) \cm_5^{(0)} \qquad \propto \, V_3 \sim \mathcal{L}^{[3,{\rm M}]} \nonumber \\
&& \,\,\,+  \sum_{k=2,4} \left[ \ck_u^{(k)} +(\mbox{tr} \gamma_9^k)^2 \ca_{99}^{(k)} +  (\mbox{tr} \gamma_9^{2k})  \cm_9^{(k)}\right] 
\, \qquad \qquad \qquad \qquad   \propto \, V_2 \sim \mathcal{L}^{[2,{\rm M}]}  \nonumber \\
&& \,\,\,+  \sum_{k=1,5} \left[ \ck_u^{(k)} +(\mbox{tr} \gamma_5^{2k})  \cm_5^{(k)}\right] +
\sum_{k=2,4} (\mbox{tr} \gamma_5^k)^2 \ca_{55}^{(k)}  
\qquad \qquad \qquad   \propto \, \frac{1}{V_2} \sim \mathcal{L}^{[2,{\rm W}]}\ , \label{sector-sum-N=2}
\eea
where from \eqref{D} we have
\be
D_\sigma^{(k)} &=&(-1)^{\gamma_3} f(\gamma_1) f(\gamma_2)    \quad \qquad\qquad  \textrm{for $\cm_9^{(3)}$, $\cm_5^{(0)}$} \\
D_\sigma^{(k)} &=&(-1)^{\gamma_2- \frac{1}{2}} (-2 \sin\pi\gamma_3) f(\gamma_1)     \qquad    \textrm{for $\cm_5^{(1,5)}$} \\
D_\sigma^{(k)} &=&1 \qquad  \qquad \qquad \qquad \qquad \qquad   \textrm{for $\ck_t^{(0,3)}$, $\ca_{95}^{(0)}$} \\
D_\sigma^{(k)} &=& (-1)^{\gamma_2} (-2 \sin\pi\gamma_3) f(\gamma_1) 
\qquad \textrm{for $\ck_u^{(1,2,4,5)}$, $\ca_{99}^{(2,4)}$, $\ca_{55}^{(2,4)}$, $\cm_9^{(2,4)}$}\ . 
\ee 
In \eqref{2ndeq} we used the tadpole condition $\mbox{tr}(\gamma_9^k)=0=\mbox{tr}(\gamma_5^k)$ for $k=3$ \cite{Aldazabal:1998mr}. Each line of \eqref{sector-sum-N=2} is proportional to a different volume factor as shown to the right, cf.\ table \ref{tab:Vol}. The first line of \eqref{sector-sum-N=2} is the analog of the contributions appearing in the $\mathbb{T}^2 \times \mathbb{T}^4 / \mathbb{Z}_2$ example discussed in \cite{Antoniadis:1996vw}. The second and third lines of \eqref{sector-sum-N=2} illustrate the general discussion below eq.\ \eqref{L_W}: the momentum sum along the second torus arises from D9-branes (parallel to the second torus) whereas the winding sum arises from D5-branes (transversal to the second torus). Using \eqref{sigma^k-2} and \eqref{D} together with \reftab{tab:details} and $\tilde{\chi}(\Theta^3,\Theta^0)=16=\tilde{\chi}(\Theta^3,\Theta^3)$ for $\K$ and $\mbox{tr}(\gamma_9^0)=32=\mbox{tr}(\gamma_5^0)$, we have
\be
\sum_{k=0,3} \ck_t^{(k)} &=& 128 \pi\, \mathcal{L}^{[3,{\rm M}]}  \label{L_M3} \\
(\mbox{tr} \gamma_9^0) (\mbox{tr} \gamma_5^0) \ca_{95}^{(0)}  &=& 16\cdot128 \pi \, \mathcal{L}^{[3,{\rm M}]}\\
(\mbox{tr} \gamma_9^{6}) \cm_9^{(3)}+(\mbox{tr} \gamma_5^{0}) \cm_5^{(0)} &=& (-2) \cdot128 \pi\, \mathcal{L}^{[3,{\rm M}]}
\ee  
for the first line of \eqref{sector-sum-N=2},
\be
\sum_{k=2,4} \ck_u^{(k)} &=& -24 \pi\, \mathcal{L}^{[2,{\rm M}]} \\
\sum_{k=2,4} (\mbox{tr} \gamma_9^k)^2 \ca_{99}^{(k)} &=& - 16\cdot24 \pi\, \mathcal{L}^{[2,{\rm M}]} \\
\sum_{k=2,4} (\mbox{tr} \gamma_9^{2k}) \cm_{9}^{(k)} &=&  2\cdot24 \pi\, \mathcal{L}^{[2,{\rm M}]}
\ee  
for the second line and
\be
\sum_{k=1,5} \ck_u^{(k)} &=& -24 \pi\, \mathcal{L}^{[2,{\rm W}]} \\
\sum_{k=2,4} (\mbox{tr} \gamma_5^k)^2 \ca_{55}^{(k)} &=& - 16\cdot24 \pi\, \mathcal{L}^{[2,{\rm W}]} \\
\sum_{k=1,5} (\mbox{tr} \gamma_5^{2k}) \cm_{5}^{(k)} &=&  2\cdot24 \pi\, \mathcal{L}^{[2,{\rm W}]}  \label{L_W2}
\ee  
for the third line.

In the above expressions we separated the relative factors so as to see the UV-divergence cancellation more easily. Concretely, from  \eqref{sector-sum-N=2}, \eqref{L_M3}-\eqref{L_W2}, \eqref{t-integral-L} and \eqref{Gammadef} we obtain
\be
&&\sum_{\sigma} \int_{\frac{1}{e_\sigma \Lambda}}^{\infty} \frac{\dd t}{t^2} \sum_{k \in \{\N=2 \}} {\rm CP}_\sigma \, \sigma^{(k)} \nonumber\\
&& =\frac{\pi\, \Lambda^2}{2} \,\Big(e_\ck^2+16 e_\ca^2 - 2 e_\cm^2\Big) \Big(128\, C^{[3,{\rm M}]}- 24\, C^{[2,{\rm M}]}-24\, C^{[2,{\rm W}]} \Big)
\nonumber \\
&& \quad + 120 \pi\, \Big(16\, C^{[3,{\rm M}]} \,\Gamma^{[3,{\rm M}]}- 3\, C^{[2,{\rm M}]} \,\Gamma^{[2,{\rm M}]}
-3\, C^{[2,{\rm W}]}\,\Gamma^{[2,{\rm W}]}\Big) \\
&& = 120 \pi \,\Big(16\, C^{[3,{\rm M}]}\, \Gamma^{[3,{\rm M}]}- 3\, C^{[2,{\rm M}]}\, \Gamma^{[2,{\rm M}]}-3\, C^{[2,{\rm W}]}\,\Gamma^{[2,{\rm W}]}\Big)\ .
\label{finite-2}
\ee
In the second equality we used $e_\ck^2+16 e_\ca^2 - 2 e_\cm^2=0$, cf.\ \eqref{e_sigma}. Thus the UV-divergences cancel. For the finite piece we obtain, using \eqref{delta-2}, \eqref{finite-2}, \eqref{C_M-W} and \eqref{Gamma_E},
\be
(\delta E)^{(\N=2)} &=& - \frac{15 \pi^2 (\alpha')^2}{8 N (4 \pi^2 \alpha')^2} \,
\,\Big(16\, C^{[3,{\rm M}]}\, \Gamma^{[3,{\rm M}]}- 3\, C^{[2,{\rm M}]}\, \Gamma^{[2,{\rm M}]}-3\, C^{[2,{\rm W}]}\,\Gamma^{[2,{\rm W}]}\Big) \\
&=&- \frac{5}{256 \pi^2} \, \left(
\frac{4\, V_3}{ \pi^2 \alpha'}\, \Gamma^{[3,{\rm M}]}- \frac{ 3\, V_2}{4 \pi^2 \alpha'}\, \Gamma^{[2,{\rm M}]}-
\frac{12 \pi^2 \alpha'}{V_2}\,\Gamma^{[2,{\rm W}]}
\right) \\
&=&-\frac{5}{256 \pi^2} \, \left[
\frac{64 \pi^2 \alpha' }{ V_3}\, E_2\big(U^{[3]} \big)- \frac{ 12 \pi^2 \alpha'}{V_2}\, E_2\big(U^{[2]}\big) - 
 \frac{3 V_2}{ 4 \pi^2 \alpha'}\,E_2\left(-\frac{1}{U^{[2]}}\right)
\right] .
\ee

Altogether, adding up the $\cn =1$ and $\cn = 2$ contributions from $\K, \A$ and $\M$ and also the contributions from $\ct$ and $S_2$ given in \eqref{deltaETS}, we obtain
\be
\delta E &=& \frac{5}{64 \pi^2}\, {\rm Cl}_2\left(\frac{\pi}{3}\right) - \frac{5}{256 \pi^2} \, \left[
\frac{64 \pi^2 \alpha' }{ V_3}\, E_2\big(U^{[3]}\big)- \frac{ 12 \pi^2 \alpha'}{V_2}\, E_2\big(U^{[2]}\big) - 
\frac{3 V_2}{ 4 \pi^2 \alpha'}\,E_2\left(-\frac{1}{U^{[2]}}\right) \right] \nonumber \\
&& + \frac{\chi}{(2 \pi)^3} \Big( 2 \zeta (3) \frac{e^{-2 \Phi_4}}{{\cal V}} + \frac{\pi^2}{3} \Big)\ , \label{finaldE}
\ee
where ${\rm Cl}_2 (\pi/3) \approx 1.015$ and the Euler number of the $\mathbb{Z}_6'$ orientifold is $\chi = 2 (h^{(1,1)}-h^{(2,1)}) = 48$, cf.\ table 2 in \cite{Klein:2000hf}, for instance. Note the term proportional to $V_2$ in \eqref{finaldE} which survives the large volume limit. It can be traced back to the contribution of winding modes, cf.\ the last line of \eqref{sector-sum-N=2}. Such terms (which survive the large volume limit) were absent in the $\cn = 2$ model discussed in \cite{Antoniadis:1996vw}, but a similar term was found by \cite{Epple:2004ra} in an $\cn = 1$ model in type IIA. At first sight it might be a bit surprising that it is the contribution of the winding modes that survives the large volume limit, given that the winding states become very heavy in this limit. However, this intuition has to be utilized with care in cases where one has an infinite tower of winding states. In that case the contribution of the winding modes in the open string channel can be reinterpreted via a Poisson resummation as arising from KK momentum modes in the closed string channel.


\section{Conclusions and Outlook}
\label{sec:concl}

We determined the quantum corrections to the Einstein-Hilbert term in toroidal minimally supersymmetric type IIB orientifolds at 1-loop order. The contributions from annulus, M\"obius and Klein Bottle are given by the general formula \eqref{1pt}, which is very similar to the formula derived in \cite{Epple:2004ra} for type IIA orientifolds with D6-branes at angles. We then evaluated this formula in concrete examples (the $\mathbb{Z}_3$ and $\mathbb{Z}_6'$ models). In doing so we encountered a new type of contributions which was absent in \cite{Epple:2004ra}. It arises from the annulus with one end on a D9-brane and one on a D5-brane, as well as from the twisted Klein Bottle. We found non-trivial contributions both from the $\cn=2$ and the $\cn = 1$ sectors of the annulus, M\"obius and Klein Bottle amplitudes (as usual, $\cn=4$ sectors do not contribute). This is in contrast to the result of \cite{Kohlprath:2003pu} which only found contributions from the torus (for orbifolds of odd order). Moreover, the resulting correction to the Einstein-Hilbert term from the $\cn = 2$ sectors in the $\mathbb{Z}_6'$ model has the interesting feature that it does not vanish in the limit of large internal volume, cf.\ the term proportional to $V_2$ in \eqref{finaldE}. This is different from the $\cn = 2$ case discussed in \cite{Antoniadis:1996vw} and similar to the situation in minimally supersymmetric type IIA toroidal orientifolds discussed in \cite{Epple:2004ra}. 

Our main motivation to consider 1-loop corrections to the Einstein-Hilbert term is their importance for determining the 1-loop corrections to the K\"ahler potential of the moduli, cf.\ \eqref{finalmetric}. A complete determination of these corrections also requires a knowledge of the correct definition of the field variables at 1-loop level. One strategy to obtain the quantum corrections to $\ImT$, cf.\ \eqref{deltaImT}, would be to use that the D5-brane gauge coupling is the imaginary part of a holomorphic function of the moduli fields. At leading order it is given by $\tau^{(0)}$. This arises at disk level. Any correction of order $e^{2 \Phi_{10}}$ relative to this (i.e.\ any 1-loop correction to this) would have to arise from genus-3/2 contributions to the gauge coupling. It would be interesting to determine these following the preliminary work of \cite{Blau:1987pn,Bianchi:1988fr,Bianchi:1989du,Antoniadis:2004qn}. We leave this for future work.


\section{Acknowledgments}
We thank Sebastian Braig for collaboration at an early stage of the project, Stefan Hofmann for interesting discussions, Marcus Berg for helpful comments on the manuscript and Michele Cicoli for pointing out a missing factor of $\alpha'$ in the first version of the paper. 
The work of M.H. is supported by the Excellence Cluster ``The Origin and the Structure of the Universe'' in Munich, by the German Research Foundation (DFG) within the Emmy-Noether-Program (grant number: HA 3448/3-1) and by the GIF under grant number 1156/2011.
The work of J.K. has been supported in parts by the Jiangsu Ministry of Science and Technology under contract BK20131264, Natural Science Foundation of China (Project No. 11405084), the Fundamental Research Funds for
the Central Universities (Project No. 020414340080) and the visitor program of the Kavli Institute for Theoretical Physics China (KITPC) in Beijing.  J.K. also acknowledges the Priority Academic Program Development for Jiangsu Higher Education Institutions (PAPD). 

\appendix

\section{Useful formulas}
\label{useful}
The $\tht$ functions are
\be 
\tht\ba{\vec \alpha}{\vec\beta}(\vec \nu,G) &=& \sum_{\vec n\in \mathbb{Z}^N} 
 e^{i\pi(\vec n+\vec \alpha)^{\rm T} G (\vec n+\vec \alpha)} 
e^{2\pi i(\vec \nu+\vec \beta)^{\rm T}(\vec n+\vec\alpha)}  \ .
\label{thetamatrix}
\ee
Poisson resummation:
\beqn \label{Poisson}
\tht\ba{\vec 0}{\vec 0}(0,it G^{-1}) &=& \sqrt{G} \, t^{-N/2}\,  
 \tht\ba{\vec 0}{\vec 0 }(0,it^{-1} G) \ .
\eeqn
Modular transformation for annulus and Klein bottle:
\be \label{modtransannulus}
\thba{\alpha}{\beta}(\nu,\tau)
&=& (-i \tau)^{-1/2} e^{2\pi i \alpha \beta - \pi i \nu^2/\tau}
\thba{-\beta}{\alpha}(\nu/\tau,-1/\tau) \label{S} \ . 
\ee
Shifts in characteristics:
\be \label{shift}
\thba{\alpha+1}{\beta}(\nu,\tau) &=& \thba{\alpha}{\beta}(\nu,\tau)\ , \non
\thba{\alpha}{\beta+1}(\nu,\tau) &=& e^{2 \pi i \alpha} \thba{\alpha}{\beta}(\nu,\tau)\ .
\ee 
$\nu$-periodicity formula:
\be  \label{period}
\thba{\alpha}{\beta}(\nu+a\tau + b,\tau)
&=&
e^{-2\pi i ab} e^{-\pi i a^2\tau}e^{-2\pi i a(\nu+\beta)}
\thba{\alpha+a}{\beta+b}(\nu,\tau)\ . 
\ee


\section{The partition function of $\mathbb{Z}_6'$}
\label{partfun}

The partition function for the world-sheet $\sigma$ reads
\begin{equation}
\langle1\rangle_\sigma=Z_\sigma= \frac{V_4}{8N (4 \pi^2 \alpha')^2}  \int_0^\infty \frac{d t}{t^3} \sum_{k=0}^{N-1} \,
\sum_{s=\text{even}} Z_\sigma^{(k)}(\tau_\sigma, s)
\end{equation}         
with
\be
Z_\A^{(k)}(\tau_\A, s) &=&  Z_{99}^{(k)}(\tau_\A, s)+ Z_{55}^{(k)}(\tau_\A, s)+Z_{95}^{(k)}(\tau_\A, s)\ ,  \\
Z_\M^{(k)}(\tau_\M, s) &=& Z_{9}^{(k)}(\tau_\M, s)+ Z_{5}^{(k)}(\tau_\M, s)\ ,  \\
Z_\K^{(k)}(\tau_\K, s) &=&  Z_{u}^{(k)}(\tau_\K, s)+ Z_{t}^{(k)}(\tau_\K, s)\ ,
\ee
where the indices $u$ and $t$ for the Klein bottle indicate the contributions of untwisted and $\Theta^3$-twisted strings running in the loop, cf.\ the caption of table \ref{tab:details}. 
In the following, exemplarily we list the full expressions for $\A$, $\M$ and $\K$ for the case of $\mathbb{Z}_6'$. In this appendix we do not assume that all the D5-branes sit at the origin of the transverse space. Rather we allow them to sit at different fixed points. The partition function for this $\mathbb{Z}_6'$ orientifold is, of course, in principle already contained in \cite{Aldazabal:1998mr} (for a partial list, see also app.\ B of \cite{Antoniadis:1999ge}). We nevertheless hope that the complete list below is useful, as the presentation slightly differs from the one given in \cite{Aldazabal:1998mr}. First we give all the different amplitudes {\it before} the spin-structure sum and second, for the M\"obius amplitude we use the half-shifted torus parameter $\tau_\M$ in the $\vartheta$-functions.


\subsection{Annulus}

\be
Z_{99}^{(k=0)}(\tau_\A)&=&\frac{ V_1 V_2 V_3}{ (4\pi^2\alpha't)^3} \big(\mathrm{tr}\gamma_{0,9}\big)^2 \times \nonumber \\
&& \times \left( \sum_{\alpha,\beta=0,\frac{1}{2}} \eta_{\alpha \beta} \prod_{i=0}^3 \frac{\vt\ss{\alpha}{\beta}}{\eta^3} \right)
\prod_{j=1}^{3} \sum_{m^1,m^2} e^{-\frac{\pi}{t} m^a m^b g_{ab}^{[j]}}\ ,
\ee
\be
Z_{99}^{(k=1,5)}(\tau_\A)&=& \big(\mathrm{tr}\gamma_{k,9}\big)^2
\prod_{j=1}^3 (-2 \sin \pi k v_j) \times \nonumber \\
&& \times \left( \sum_{\alpha,\beta=0,\frac{1}{2}} \eta_{\alpha \beta}  \frac{\vt\ss{\alpha}{\beta}}{\eta^3} 
\prod_{i= 1}^3 \frac{ \vt\ss{\alpha}{\beta+k v_i} }{\vt\ss{1/2}{1/2+k v_i}} \right),
\ee
\be
Z_{99}^{(k=2,4)}(\tau_\A)&=&\frac{ V_2}{(4 \pi^2 \alpha' t)} \big(\mathrm{tr}\gamma_{k,9}\big)^2 
 \times \nonumber \\
&& \times \prod_{j=1,3} (-2 \sin \pi k v_j) \sum_{m^1,m^2} e^{-\frac{\pi}{t} m^a m^b g_{ab}^{[2]}} \nonumber \\
&& \times \left( \sum_{\alpha,\beta=0,\frac{1}{2}} \eta_{\alpha \beta}  \frac{\vt\ss{\alpha}{\beta}}{\eta^3} \frac{\vt\ss{\alpha}{\beta+k v_2}}{\eta^3} 
\prod_{i= 1,3} \frac{ \vt\ss{\alpha}{\beta+k v_i} }{\vt\ss{1/2}{1/2+k v_i}} \right),
\ee
\be
Z_{99}^{(k=3)}(\tau_\A)&=&\frac{ V_3}{(4 \pi^2 \alpha' t)} \big(\mathrm{tr}\gamma_{3,9}\big)^2 
 \sum_{m^1,m^2} e^{-\frac{\pi}{t} m^a m^b g_{ab}^{[3]}}  	\times \nonumber \\
&& \times \left( \sum_{\alpha,\beta=0,\frac{1}{2}} \eta_{\alpha \beta}  \frac{\vt\ss{\alpha}{\beta}}{\eta^3} \frac{\vt\ss{\alpha}{\beta+k v_3}}{\eta^3} 
\prod_{i= 1,2} \frac{2 \vt\ss{\alpha}{\beta+k v_i} }{\vt\ss{1/2}{0}} \right),
\ee
\be
Z_{55}^{(k=0)}(\tau_\A) &=& \frac{ V_3 (4 \pi^2 \alpha')^2}{ V_1 V_2 \, t^2 (4 \pi^2 \alpha' t) }  
(\mathrm{tr} \gamma_{0,5})^2
\prod_{j=1}^2 \sum_{w_1,w_2} e^{-\frac{\pi}{ t} w_a w_b g^{[j]ab}} \times \nonumber \\
&& \times \sum_{m^1,m^2} e^{-\frac{\pi}{ \alpha'  t} m^a m^b g_{ab}^{[3]}}
 \left( \sum_{\alpha,\beta=0,\frac{1}{2}} \eta_{\alpha \beta} \prod_{i=0}^3 \frac{\vt\ss{\alpha}{\beta}}{\eta^3} 
 \right),
\ee
\be
Z_{55}^{(k=1,5)}(\tau_\A)&=& \sum_{L=0}^3 \big(\mathrm{tr}\gamma_{k,5,L}\big)^2
\prod_{j=1}^3 (-2 \sin \pi k v_j) \times \nonumber \\
&& \times \left( \sum_{\alpha,\beta=0,\frac{1}{2}} \eta_{\alpha \beta}  \frac{\vt\ss{\alpha}{\beta}}{\eta^3} 
\prod_{i= 1}^3 \frac{ \vt\ss{\alpha}{\beta+k v_i} }{\vt\ss{1/2}{1/2+k v_i}} \right),
\ee
\be
Z_{55}^{(k=2,4)}(\tau_\A)&=&\frac{ (4 \pi^2 \alpha') }{ V_2  \, t} \sum_{M=0}^2 \big(\mathrm{tr}\gamma_{k,5,M}\big)^2 
 \times \nonumber \\
&& \times \prod_{j=1,3} (-2 \sin \pi k v_j) \sum_{w_1,w_2} e^{-\frac{\pi}{  t} w_a w_b g^{[2]ab}} \nonumber \\
&& \times \left( \sum_{\alpha,\beta=0,\frac{1}{2}} \eta_{\alpha \beta}  \frac{\vt\ss{\alpha}{\beta}}{\eta^3} \frac{\vt\ss{\alpha}{\beta+k v_2}}{\eta^3} 
\prod_{i= 1,3} \frac{ \vt\ss{\alpha}{\beta+k v_i} }{\vt\ss{1/2}{1/2+k v_i}} \right),
\ee
\be
Z_{55}^{(k=3)}(\tau_\A)&=&\frac{ V_3}{(4 \pi^2 \alpha' t)} \sum_{I=0}^{15} \big(\mathrm{tr}\gamma_{3,5,I}\big)^2 
 \sum_{m^1,m^2} e^{-\frac{\pi}{t} m^a m^b g_{ab}^{[3]}}  	\times \nonumber \\
&& \times \left( \sum_{\alpha,\beta=0,\frac{1}{2}} \eta_{\alpha \beta}  \frac{\vt\ss{\alpha}{\beta}}{\eta^3} \frac{\vt\ss{\alpha}{\beta+k v_3}}{\eta^3} 
\prod_{i= 1,2} \frac{2 \vt\ss{\alpha}{\beta+k v_i} }{\vt\ss{1/2}{0}} \right),
\ee
\be
Z_{95}^{(k=0)}(\tau_\A)&=&\frac{ V_3}{(4 \pi^2 \alpha' t)} \mathrm{tr}\gamma_{0,9} \mathrm{tr}\gamma_{0,5}
 \sum_{m^1,m^2} e^{-\frac{\pi}{t} m^a m^b g_{ab}^{[3]}}  	\times \nonumber \\
&& \times \left( \sum_{\alpha,\beta=0,\frac{1}{2}} \eta_{\alpha \beta}  \left( \prod_{i=0,3}\frac{\vt\ss{\alpha}{\beta}}{\eta^3}  \right)
 \prod_{i=1}^2\frac{\vt\ss{\alpha+1/2}{\beta} }{\vt\ss{0}{1/2}} \right),
\ee
\be
Z_{95}^{(k=1,5)}(\tau_\A)&=&\sum_{L=0}^3 \mathrm{tr} \gamma_{k,9} \mathrm{tr} \gamma_{k,5,L} 
  (-2 \sin \pi k v_3) \times \nonumber \\
&& \times \left( \sum_{\alpha,\beta=0,\frac{1}{2}} \eta_{\alpha \beta}  \frac{\vt\ss{\alpha}{\beta}}{\eta^3} 
 \frac{\vt \ss{\alpha}{\beta+k v_3} }{\vt\ss{1/2}{1/2+k v_3}} \prod_{i=1}^2 \frac{\vt \ss{\alpha+1/2}{\beta+k v_i} }{\vt\ss{0}{1/2+k v_i}} 
 \right),
\ee
\be
Z_{95}^{(k=2,4)}(\tau_\A)&=& \sum_{M=0}^2 \mathrm{tr} \gamma_{k,9} \mathrm{tr}\gamma_{k,5,M} 
  (-2 \sin \pi k v_3)  \times \nonumber \\
&& \times \left( \sum_{\alpha,\beta=0,\frac{1}{2}} \eta_{\alpha \beta}  \frac{\vt\ss{\alpha}{\beta}}{\eta^3} 
 \frac{\vt \ss{\alpha}{\beta+k v_3} }{\vt\ss{1/2}{1/2+k v_3}} \prod_{i=1}^2 \frac{\vt \ss{\alpha+1/2}{\beta+k v_i} }{\vt\ss{0}{1/2+k v_i}} 
  \right),
\ee
\be
Z_{95}^{(k=3)}(\tau_\A)&=&\frac{ V_3}{(4 \pi^2 \alpha' t)} \sum_{I=0}^{15} \mathrm{tr}\gamma_{3,9} \mathrm{tr}\gamma_{3,5,I}
 \sum_{m^1,m^2} e^{-\frac{\pi}{t} m^a m^b g_{ab}^{[3]}}  	\times \nonumber \\
&& \times \left( \sum_{\alpha,\beta=0,\frac{1}{2}} \eta_{\alpha \beta}  \frac{\vt\ss{\alpha}{\beta}}{\eta^3} \frac{\vt\ss{\alpha}{\beta+k v_3}}{\eta^3}  
 \prod_{i=1}^2\frac{\vt\ss{\alpha+1/2}{\beta+k v_i} }{\vt\ss{0}{1/2+k v_i}} \right).
\ee


\subsection{M\"obius}
\be
Z_9^{(k=0)}(\tau_\M)&=&-\frac{ V_1 V_2 V_3}{ (4\pi^2\alpha't)^3} \mathrm{tr} \big((\gamma_{\Omega_0,9})^\mathrm{T}(\gamma_{\Omega_0,9})^{-1}\big) \times \nonumber \\
&& \times \left( \sum_{\alpha,\beta=0,\frac{1}{2}} \eta_{\alpha \beta} \prod_{i=0}^3 \frac{\vt\ss{\alpha}{\beta}}{\eta^3} \right)
\prod_{j=1}^{3} \sum_{m^1,m^2} e^{-\frac{\pi}{t} m^a m^b g_{ab}^{[j]}} \ ,
\ee
\be
Z_9^{(k=1,5)}(\tau_\M)&=&- \mathrm{tr} \big((\gamma_{\Omega_k,9})^\mathrm{T}(\gamma_{\Omega_k,9})^{-1}\big) 
\prod_{j=1}^3 (-2 \sin \pi k v_j) \times \nonumber \\
&& \times \left( \sum_{\alpha,\beta=0,\frac{1}{2}} \eta_{\alpha \beta}  \frac{\vt\ss{\alpha}{\beta}}{\eta^3} 
\prod_{i= 1}^3 \frac{ \vt\ss{\alpha}{\beta+k v_i} }{\vt\ss{1/2}{1/2+k v_i}} \right),
\ee
\be
Z_9^{(k=2,4)}(\tau_\M)&=&-\frac{V_2}{(4 \pi^2 \alpha' t)} \mathrm{tr} \big((\gamma_{\Omega_k,9})^\mathrm{T}(\gamma_{\Omega_k,9})^{-1}\big) 
 \times \nonumber \\
&& \times \prod_{j=1,3} (-2 \sin \pi k v_j) \sum_{m^1,m^2} e^{-\frac{\pi}{t} m^a m^b g_{ab}^{[2]}} \nonumber \\
&& \times \left( \sum_{\alpha,\beta=0,\frac{1}{2}} \eta_{\alpha \beta}  \frac{\vt\ss{\alpha}{\beta}}{\eta^3} \frac{\vt\ss{\alpha}{\beta+k v_2}}{\eta^3} 
\prod_{i= 1,3} \frac{ \vt\ss{\alpha}{\beta+k v_i} }{\vt\ss{1/2}{1/2+k v_i}} \right),
\ee
\be
Z_9^{(k=3)}(\tau_\M)&=&-\frac{ V_3}{(4 \pi^2 \alpha' t)} \mathrm{tr} \big((\gamma_{\Omega_3,9})^\mathrm{T}(\gamma_{\Omega_3,9})^{-1}\big) 
 \sum_{m^1,m^2} e^{-\frac{\pi}{t} m^a m^b g_{ab}^{[3]}}  	\times \nonumber \\
&& \times \left( \sum_{\alpha,\beta=0,\frac{1}{2}} \eta_{\alpha \beta}  \frac{\vt\ss{\alpha}{\beta}}{\eta^3} \frac{\vt\ss{\alpha}{\beta+k v_3}}{\eta^3} 
\prod_{i= 1,2} \frac{2 \vt\ss{\alpha}{\beta+k v_i} }{\vt\ss{1/2}{0}} \right),
\ee
\be
Z_5^{(k=0)}(\tau_\M)&=&-\frac{V_3}{(4 \pi^2 \alpha' t)} \mathrm{tr} \big((\gamma_{\Omega_0,5})^\mathrm{T}(\gamma_{\Omega_0,5})^{-1}\big) 
 \sum_{m^1,m^2} e^{-\frac{\pi}{t} m^a m^b g_{ab}^{[3]}}  	\times \nonumber \\
&& \times \left( \sum_{\alpha,\beta=0,\frac{1}{2}} \eta_{\alpha \beta}  \left( \prod_{i=0,3}\frac{\vt\ss{\alpha}{\beta}}{\eta^3}  \right)
 \frac{2 \vt\ss{\alpha}{\beta+1/2} }{\vt\ss{1/2}{0}}  \frac{2 \vt\ss{\alpha}{\beta-1/2} }{\vt\ss{1/2}{0}}\right),
\ee
\be
Z_5^{(k=1,5)}(\tau_\M)&=&-\frac{ (4 \pi^2 \alpha')}{ V_2 \, t} \sum_{L=0}^3 \mathrm{tr} \big((\gamma_{\Omega_k,5,L})^\mathrm{T}(\gamma_{\Omega_k,5,L})^{-1}\big) 
 \times \nonumber \\
&& \times  (-2 \sin \pi k v_3) (2 \cos \pi k v_1) \sum_{w_1,w_2} e^{-\frac{\pi}{ t} w_a w_b g^{[2]ab}} \nonumber \\
&& \times \left( \sum_{\alpha,\beta=0,\frac{1}{2}} \eta_{\alpha \beta}  \frac{\vt\ss{\alpha}{\beta}}{\eta^3} 
\frac{\vt\ss{\alpha}{\beta+k v_1 +1/2}}{\vt\ss{1/2}{kv_1}} \frac{\vt\ss{\alpha}{\beta+kv_2-1/2}}{\eta^3} 
 \frac{ \vt\ss{\alpha}{\beta+k v_3} }{\vt\ss{1/2}{1/2+k v_3}} \right),
\ee
\be
Z_5^{(k=2,4)}(\tau_\M)&=&-\sum_{M=0}^2 \mathrm{tr} \big((\gamma_{\Omega_k,5,M})^\mathrm{T}(\gamma_{\Omega_k,5,M})^{-1}\big) 
  (-2 \sin \pi k v_3) \prod_{j=1,2}(2 \cos \pi k v_j) \times \nonumber \\
&& \times \left( \sum_{\alpha,\beta=0,\frac{1}{2}} \eta_{\alpha \beta}  \frac{\vt\ss{\alpha}{\beta}}{\eta^3} 
\frac{\vt\ss{\alpha}{\beta+k v_1 +1/2}}{\vt\ss{1/2}{kv_1}} \frac{ \vt\ss{\alpha}{\beta+kv_2-1/2}}{\vt\ss{1/2}{kv_2}} 
 \frac{ \vt\ss{\alpha}{\beta+k v_3} }{\vt\ss{1/2}{1/2+k v_3}} \right),
\ee
\be
Z_5^{(k=3)}(\tau_\M)&=&-\frac{V_3 (4 \pi^2 \alpha')^2}{ (4 \pi^2 \alpha' t) V_1 V_2\, t^2} \sum_{I=0}^{15} 
\mathrm{tr} \big((\gamma_{\Omega_3,5,I})^\mathrm{T}(\gamma_{\Omega_3,5,I})^{-1}\big) 
 \times \nonumber \\
&& \times \sum_{m^1,m^2} e^{-\frac{\pi}{t} m^a m^b g_{ab}^{[3]}}  
\prod_{j=1}^2 \sum_{w_1,w_2} e^{-\frac{\pi}{  t} w_a w_b g^{[j]ab}} \nonumber \\
&& \times \left( \sum_{\alpha,\beta=0,\frac{1}{2}} \eta_{\alpha \beta}  \frac{\vt\ss{\alpha}{\beta}}{\eta^3} 
\frac{\vt\ss{\alpha}{\beta+k v_1 +1/2}}{\eta^3} \frac{\vt\ss{\alpha}{\beta+kv_2-1/2}}{\eta^3} 
 \frac{ \vt\ss{\alpha}{\beta+k v_3} }{\eta^3} \right).
\ee


\subsection{Klein Bottle}
 
\be
Z_{u}^{(k=0)}(\tau_\K)&=&\frac{ V_1 V_2 V_3}{(4\pi^2\alpha't)^3}   \prod_{j=1}^{3} \sum_{m^1,m^2} e^{-\frac{\pi}{t} m^a m^b g_{ab}^{[j]}} \times \nonumber \\
&& \times \left( \sum_{\alpha,\beta=0,\frac{1}{2}} \eta_{\alpha \beta} \prod_{i=0}^3 \frac{\vt\ss{\alpha}{\beta}}{\eta^3} \right),
\ee
\be
Z_{u}^{(k=1,5)}(\tau_\K)&=&\frac{ (4 \pi^2 \alpha') }{ V_2 \,  t} 
\prod_{j=1,3} (-2 \sin 2 \pi k v_j) \sum_{w_1,w_2} e^{-\frac{\pi}{t} w_a w_b g^{[2]ab}} \times \nonumber \\
&& \times \left( \sum_{\alpha,\beta=0,\frac{1}{2}} \eta_{\alpha \beta}  \frac{\vt\ss{\alpha}{\beta}}{\eta^3} 
\frac{\vt\ss{\alpha}{\beta+2k v_2}}{\eta^3} 
\prod_{i= 1,3} \frac{ \vt\ss{\alpha}{\beta+2k v_i} }{\vt\ss{1/2}{1/2+2k v_i}} \right),
\ee
\be
Z_{u}^{(k=2,4)}(\tau_\K)&=&\frac{V_2}{(4 \pi^2 \alpha' t)} 
 \sum_{m^1,m^2} e^{-\frac{\pi}{t} m^a m^b g_{ab}^{[2]}}  \prod_{j=1,3} (-2 \sin 2 \pi k v_j) \nonumber \\
&& \times \left( \sum_{\alpha,\beta=0,\frac{1}{2}} \eta_{\alpha \beta}  \frac{\vt\ss{\alpha}{\beta}}{\eta^3} 
\frac{\vt\ss{\alpha}{\beta+2 k v_2}}{\eta^3} 
\prod_{i= 1,3} \frac{ \vt\ss{\alpha}{\beta+2 k v_i} }{\vt\ss{1/2}{1/2+2 k v_i}} \right),
\ee
\be
Z_{u}^{(k=3)}(\tau_\K) &=& \frac{ V_3 (4 \pi^2 \alpha')^2}{ V_1 V_2 \, t^2 (4 \pi^2 \alpha' t) }  
\prod_{j=1}^2 \sum_{w_1,w_2} e^{-\frac{\pi}{ t} w_a w_b g^{[j]ab}} \times \nonumber \\
&& \times \sum_{m^1,m^2} e^{-\frac{\pi}{ \alpha'  t} m^a m^b g_{ab}^{[3]}}
 \left( \sum_{\alpha,\beta=0,\frac{1}{2}} \eta_{\alpha \beta} \frac{\vt\ss{\alpha}{\beta}}{\eta^3} 
 \prod_{i=1}^3 \frac{\vt\ss{\alpha}{\beta+2 k v_i}}{\eta^3} 
 \right),
\ee
\be
Z_{t}^{(k=0)}(\tau_\K)&=&\frac{ V_3}{(4 \pi^2 \alpha' t)} \, \tilde{\chi}(\Theta^3,\Theta^k) \,
 \sum_{m^1,m^2} e^{-\frac{\pi}{t} m^a m^b g_{ab}^{[3]}}  	\times \nonumber \\
&& \times \left( \sum_{\alpha,\beta=0,\frac{1}{2}} \eta_{\alpha \beta}  \left( \prod_{i=0,3}\frac{\vt\ss{\alpha}{\beta}}{\eta^3}  \right)
 \prod_{i=1}^2\frac{\vt\ss{\alpha+1/2}{\beta} }{\vt\ss{0}{1/2}} \right),
\ee
\be
Z_{t}^{(k=1,5)}(\tau_\K)&=& \tilde{\chi}(\Theta^3,\Theta^k) \, 
  (-2 \sin 2 \pi k v_3) \times \nonumber \\
&& \times \left( \sum_{\alpha,\beta=0,\frac{1}{2}} \eta_{\alpha \beta}  \frac{\vt\ss{\alpha}{\beta}}{\eta^3} 
 \frac{\vt\ss{\alpha}{\beta+2 k v_3} }{\vt\ss{1/2}{1/2+2 k v_3}} 
 \prod_{i=1}^2 \frac{\vt \ss{\alpha+1/2}{\beta+2 k v_i} }{\vt\ss{0}{1/2+2 k v_i}} 
 \right),
\ee
\be
Z_{t}^{(k=2,4)}(\tau_\K)&=&  \tilde{\chi}(\Theta^3,\Theta^k) \,
  (-2 \sin 2 \pi k v_3)  \times \nonumber \\
&& \times \left( \sum_{\alpha,\beta=0,\frac{1}{2}} \eta_{\alpha \beta}  \frac{\vt\ss{\alpha}{\beta}}{\eta^3} 
 \frac{\vt \ss{\alpha}{\beta+2 k v_3} }{\vt\ss{1/2}{1/2+2 k v_3}} 
 \prod_{i=1}^2 \frac{\vt \ss{\alpha+1/2}{\beta+2 k v_i} }{\vt\ss{0}{1/2+2 k v_i}} 
  \right),
\ee
\be
Z_{t}^{(k=3)}(\tau_\K)&=&\frac{ V_3}{(4 \pi^2 \alpha' t)}\, \tilde{\chi}(\Theta^3,\Theta^k) \,
 \sum_{m^1,m^2} e^{-\frac{\pi}{t} m^a m^b g_{ab}^{[3]}}  	\times \nonumber \\
&& \times \left( \sum_{\alpha,\beta=0,\frac{1}{2}} \eta_{\alpha \beta}  \frac{\vt\ss{\alpha}{\beta}}{\eta^3} 
\frac{\vt\ss{\alpha}{\beta+2 k v_3}}{\eta^3}  
 \prod_{i=1}^2\frac{\vt\ss{\alpha+1/2}{\beta+2 k v_i} }{\vt\ss{0}{1/2+2 k v_i}} \right).
\ee


\section{Two integrals} 
\label{t-integral}

\subsection{${\cal N}=1$ sector $t$-integral}
\label{t-integral_n1}

In order to evaluate the $t$-integral of $\N = 1$ sectors with $h_i \neq 0$ (i.e.\ for $\K_t$ and $\ca_{95}$, cf.\ table \ref{tab:details}) we need the integral (assuming $0<\gamma<1$)  
\be
I= \int_{\frac{1}{e_\sigma \Lambda}}^\infty \frac{\dd t}{t^2} \,
 \frac{\vartheta_4'(\gamma,\tau_\sigma )}{\vartheta_4(\gamma, \tau_\sigma)}
\ee
with $\sigma=\ck, \ca$ and $\tau_\sigma= \frac{i e_\sigma t}{2}$ ($e_\sigma$ was defined in \eqref{e_sigma}). Evaluating this integral follows very closely a similar calculation in app.\ M of \cite{Berg:2014ama}.
By modular transformation of the Jacobi theta function (using \eqref{modtransannulus}, $\tht_4=\thba{0}{1/2}$ and $\tht_2=\thba{1/2}{0}$) we have
\be \label{modul}
\frac{\tht_4'(\gamma,i e_\sigma t/2)}{\tht_4(\gamma,i e_\sigma t/2)}=-4 \pi \gamma l - 
2 i l \frac{\tht_2'(- 2 i \gamma l, 2 i l)}{\tht_2(-2 i \gamma l,2 i l)}\ ,
\ee
where $ l \equiv \frac{1}{e_\sigma t}$. 
Using the representation for $|{\rm Im}(z)|<{\rm Im}(\tau_\sigma)$ (cf.\ prob.\ 12 on p.\ 489 of \cite{WW}) 
\be \label{abram}
\frac{\tht_2'(z)}{\tht_2(z)} &=& - \pi \tan \pi z + 4 \pi \sum_{n=1}^{\infty} \frac{(-1)^n q^n}{1-q^n} \sin 2 \pi n z \\
&=&- \pi \tan \pi z + 4 \pi \sum_{n,m=1}^{\infty} (-1)^n q^{n m} \sin 2 \pi n z\ ,
\ee 
we arrive at
\be \label{treeA}
I&=&\int_{\frac{1}{e_\sigma \Lambda}}^\infty \frac{\dd t}{t^2} \,
 \frac{\vartheta_4'(\gamma,\tau_\sigma )}{\vartheta_4(\gamma, \tau_\sigma)}  \\
&=&e_\sigma \int_0^\Lambda \dd l \left(-4 \pi \gamma l - 
2 i l \frac{\tht_2'(- 2 i \gamma l, 2 i l)}{\tht_2(-2 i \gamma l,2 i l)} \right)    \\
&=&- 2 \pi e_\sigma \int_{0}^{\Lambda} dl\,l\,\Big( 2\gamma-\tanh(2\pi \gamma l)
+ 4 \sum_{n,m=1}^{\infty} (-1)^n e^{-4 \pi l n m} \sinh(4 \pi n \gamma l)\Big) \ . \label{treeA2}
\ee
Let us start with the last term, which is free of UV divergencies (so we can set $\Lambda=\infty$):
\be
I_1 &=& - 8 \pi e_\sigma \int_{0}^{\infty} dl\,l\,  \sum_{n,m=1}^{\infty} (-1)^n e^{-4 \pi l n m} \sinh(4 \pi n \gamma l) \nonumber \\
&=& - \pi e_\sigma \sum_{n,m=1}^\infty \frac{(-1)^n \gamma m}{(\gamma^2-m^2)^2 n^2 \pi^2} \nonumber \\
&=&  - \pi e_\sigma \left(\sum_{n=1}^\infty \frac{(-1)^n}{n^2 \pi^2} \right)
 \left(\sum_{m=1}^\infty \frac{\gamma m}{(\gamma^2-m^2)^2} \right) \nonumber \\
 &=& - e_\sigma \frac{\pi}{48}[\psi'(1+\gamma)-\psi'(1-\gamma)] \nonumber \\
&=& - e_\sigma \frac{\pi}{48} \left[\psi'(\gamma)-\psi'(1-\gamma)-\frac{1}{\gamma^2}\right].
\ee
Here $\psi'(x)$ denotes the trigamma function and in the last line we used $\psi'(1+\gamma)=\psi'(\gamma)-1/\gamma^2$.

Now let us look at the first and second term in \eqref{treeA2}:
\be
I_2= - 2 \pi e_\sigma \int_{0}^{\Lambda} dl \,l (2 \gamma) = - 2 \pi e_\sigma \gamma \Lambda^2\ ,
\ee 
\be
I_3=2\pi e_\sigma \int_{0}^{\Lambda} dl \,l\,\tanh(2\pi \gamma l)
&=&e_\sigma \left[-\frac{\pi}{48 \gamma^2}+\pi \Lambda^2+\frac{\Lambda \log(1+e^{-4\gamma\Lambda\pi})}{\gamma}
-\frac{Li_2(-e^{-4\gamma\Lambda\pi})}{4 \gamma^2 \pi} \right] \nonumber \\
&\stackrel{\Lambda\to\infty}{=}&  e_\sigma\left[- \frac{\pi}{48 \gamma^2}+ \pi \Lambda^2\right].
\ee
Here $Li_2(z)$ is the dilogarithm function. In the second equality we used that the third and last term vanish as $\Lambda\to\infty$. 

In total we obtain
\be 
\int_{1\over {e_\sigma \Lambda}}^\infty \frac{dt}{t^2}  \, \frac{\tht_4'(\gamma,e_\sigma i t/2)}{\tht_4(\gamma, e_\sigma i t/2)}= I_1+I_2+I_3 = 
e_\sigma \pi(1-2\gamma) \Lambda^2 - e_\sigma \frac{\pi}{48} \left[ \psi'(\gamma) - \psi'(1-\gamma) \right] .
\ee


\subsection{${\cal N}=2$ sector $t$-integral}
\label{t-integral_n2}

The $t$-integrals appearing in $\N=2$ sectors are very similar to those determining the $\N=2$ sector corrections to the K\"ahler metric calculated in \cite{Berg:2005ja}. Concretely, they are given by 
\be
\Gamma^{[n,{\rm M/W}]} &=& \int_{0}^{\infty} \frac{\dd t}{t^3}  \sum_{\vec{m} \in \mathbb{Z}^2 \setminus \vec{0}}  
e^{- \frac{\pi}{t} m^a m^b\, g_{ab}^{[n,{\rm M/W}]}}  \\
&=& \sum_{\vec{m} \in \mathbb{Z}^2 \setminus \vec{0}}\, \int_{0}^{\infty} \frac{\dd t}{t^3}e^{- \frac{\pi}{t} m^a m^b\, g_{ab}^{[n,{\rm M/W}]}} \\
&=& \frac{1}{\pi^2} \sum_{\vec{m} \in \mathbb{Z}^2 \setminus \vec{0}} \frac{1}{\Big(m^a m^b \, g_{ab}^{[n,{\rm M/W}]} \Big)^2 }\ . \label{Gamma}
\ee
The metric $g_{ab}^{[n,{\rm M/W}]}$ is given by \eqref{g_M-W}. Using \eqref{g_M-W} and the expression for $g_{ab}^{[n]}$ in terms of the complex structure $U^{[n]}= U_1^{[n]} + i \, U_2^{[n]}$ of $n$th torus, i.e.
\be \label{torusmetric}
g_{ab}^{[n]}=  \frac{\sqrt{{\rm det} g^{[n]}}}{U_2^{[n]}} \begin{pmatrix}
  1 & U_1^{[n]} \\
  U_1^{[n]} & \big|U^{[n]}\big|^2  
  \end{pmatrix},
\ee
one can write 
\be
g_{ab}^{[n,{\rm M/W}]}= \begin{cases}
\frac{\sqrt{{\rm det} g^{[n]}}}{U_2^{[n]}}        
\begin{pmatrix}
  1 & U_1^{[n]} \\
  U_1^{[n]} & \big|U^{[n]}\big|^2  
  \end{pmatrix}
& \quad \text{for M (momentum sum)} \\
\frac{1}{\tilde{U}_2^{[n]} \sqrt{{\rm det} g^{[n]}}}        
\begin{pmatrix}
  1 & \tilde{U}_1^{[n]} \\
  \tilde{U}_1^{[n]} & \big|\tilde{U}^{[n]}\big|^2  
  \end{pmatrix}
     & \quad \text{for W (winding sum)} \\
    \end{cases}  \label{g_M-W_U}
\ee
with $\tilde{U}^{[n]} = \tilde{U}_1^{[n]}+i \, \tilde{U}_2^{[n]} = -(U^{[n]})^{-1}$ (i.e. $\tilde{U}_1^{[n]}= -U_1^{[n]} / \big|U^{[n]}\big|^{2}$ and $\tilde{U}_2^{[n]}= U_2^{[n]}/ \big|U^{[n]}\big|^{2}$). $\tilde{U}^{[n]} = -(U^{[n]})^{-1}$ follows from the fact that $g_{ab}^{[n, {\rm W}]}$ is 
the inverse matrix of $g_{ab}^{[n, {\rm M}]}$.

Then we obtain
\be
 \sum_{\vec{m} \in \mathbb{Z}^2 \setminus \vec{0}}  \frac{1}{\Big(m^a m^b \, g_{ab}^{[n, {\rm M/W}]} \Big)^2 }
= \begin{cases}
\frac{1}{{\rm det} g^{[n]}}\,  E_2\big(U^{[n]}\big),  & \quad \text{for M (momentum sum)}\\ \\
{\rm det} g^{[n]}\, E_2\big(- (U^{[n]})^{-1}\big) & \quad \text{for W (winding sum)} \ .\\
\end{cases}
\ee
Here $E_s(U)$ is the non-holomorphic Eisenstein series
\be \label{eisenstein}
E_s(U)=  \sum_{\vec{m} \in \mathbb{Z}^2 \setminus \vec{0}} \frac{U_2^s}{|m^1+m^2\, U|^{2 s}} \ .
\ee
Therefore, from \eqref{Gamma} and using $\sqrt{{\rm det} g^{[n]}} = \frac {V_n}{4 \pi^2 \alpha'}$, we obtain
\be
\Gamma^{[n,{\rm M/W}]} =  \begin{cases}
\frac{(4 \pi^2 \alpha')^2 }{\pi^2 V_n^2}\,  E_2\big(U^{[n]}\big),  & \quad \text{for M (momentum sum)}\\ \\
\frac{V_n^2}{\pi^2 ( 4 \pi^2 \alpha')^2 }\, E_2\big(- (U^{[n]})^{-1}\big) & \quad \text{for W (winding sum)}\ . \\ 
\end{cases} \label{Gamma_E_2}
\ee


\end{document}